# The role of mergers in driving morphological transformation over cosmic time

G. Martin,[1][*] S. Kaviraj,[1] J. E. G. Devriendt,[2] Y. Dubois[3] and C. Pichon[3,4]

[1]Centre for Astrophysics Research, School of Physics, Astronomy and Mathematics, University of Hertfordshire, College Lane, Hatfield AL10 9AB, UK
[2]Dept of Physics, University of Oxford, Keble Road, Oxford OX1 3RH UK
[3]Institut d'Astrophysique de Paris, Sorbonne Universités, UMPC Univ Paris 06 et CNRS, UMP 7095, 98 bis bd Arago, 75014 Paris, France
[4]Korea Institute of Advanced Studies (KIAS) 85 Hoegiro, Dongdaemun-gu, Seoul, 02455, Republic of Korea

21 July 2018

**ABSTRACT**

Understanding the processes that trigger morphological transformation is central to understanding how and why the Universe transitions from being disc-dominated at early epochs to having the morphological mix that is observed today. We use Horizon-AGN, a cosmological hydrodynamical simulation, to perform a comprehensive study of the processes that drive morphological change in massive ($M_\star/\mathrm{M_\odot} > 10^{10}$) galaxies over cosmic time. We show that (1) essentially all the morphological evolution in galaxies that are spheroids at $z = 0$ is driven by mergers with mass ratios greater than $1 : 10$, (2) major mergers alone cannot produce today's spheroid population – minor mergers are responsible for a third of all morphological transformation over cosmic time and are its dominant driver after $z \sim 1$, (3) prograde mergers trigger milder morphological transformation than retrograde mergers – while both types of events produce similar morphological changes at $z > 2$, the average change due to retrograde mergers is around twice that due to their prograde counterparts at $z \sim 0$, (4) remnant morphology depends strongly on the gas fraction of a merger, with gas-rich mergers routinely re-growing discs, and (5) at a given stellar mass, discs do not exhibit drastically different merger histories from spheroids – disc survival in mergers is driven by acquisition of cold gas (via cosmological accretion and gas-rich interactions) and a preponderance of prograde mergers in their merger histories.

## 1 INTRODUCTION

As predicted by hierarchical structure formation scenarios (e.g. Fall & Efstathiou 1980; van den Bosch et al. 2002; Agertz et al. 2011), high-redshift observations of massive galaxies indicate that the early Universe was dominated by systems possessing disc-like morphologies (e.g. Buitrago et al. 2014; Shibuya et al. 2015). In contrast, the morphological mix of today's Universe is dominated by massive galaxies with spheroidal morphologies (e.g. Bernardi et al. 2003; Conselice et al. 2014), with a majority of objects at low redshift hosting significant bulge components (e.g Lintott et al. 2011). This disparity is evidence for significant structural evolution in the galaxy population over cosmic time, as a result of which discy, rotationally-supported galaxies are steadily transformed into spheroidal, dispersion-supported systems (e.g Butcher & Oemler 1984; Dressler et al. 1997; Postman et al. 2005; Smith et al. 2005; Conselice et al. 2008; Buitrago et al. 2014). Understanding the processes that drive this morphological transformation is, therefore, central to our comprehension of how galaxies have evolved over the lifetime of the Universe.

While empirical morphological classification schemes (e.g. Hubble 1936; Cappellari et al. 2011) are largely defined using only visual or kinematic criteria, the morphological type of galaxies at the present-day is strongly aligned with their physical properties. Stellar masses, star-formation rates, colours, merger histories and local environment (e.g. Dressler 1980; Dressler et al. 1997; Strateva et al. 2001; Hogg et al. 2002; Bundy et al. 2005; Conselice 2006; Skibba et al. 2009; Bluck et al. 2014; Smethurst et al. 2015; Whitaker et al. 2015) all correlate strongly with galaxy morphology. This points towards a picture of galaxy morphology that does not depend on a single mechanism for morphological change, at least across a broad range of masses and environments.

However, disentangling the role of different mechanisms in triggering morphological change remains difficult. Many processes are likely to be involved in the transformation of discs to spheroids, and the relative contribution of these processes is not well understood. For example, the theoretical literature has long highlighted the role of mergers in the creation of spheroidal systems, as the gravitational torques can remove stars from ordered rotational orbits in discy progenitors to chaotic orbits that form dispersion-supported spheroidal remnants. Major mergers i.e. those that involve progenitors with roughly equal mass, are considered to be particularly efficient at producing spheroidal systems (e.g. Toomre 1977; Negroponte & White 1983; Di Matteo et al. 2007; Hopkins et al. 2009a; Ferreras et al. 2009; Conselice et al. 2009; Taranu et al. 2013; Naab et al. 2014; Deeley et al. 2017), although minor mergers (i.e. those with unequal progenitor mass ratios) are also likely to play a role in the transformation of morphologies, by either producing chaotic stellar orbits as major mergers do, or by triggering disc instabilities (e.g. Dekel et al. 2009; Fiacconi et al. 2015; Zolotov et al. 2015; Welker et al. 2017). Processes other than galaxy mergers may also play a role in inducing morphological transformation. For example, in very dense environments, fly-bys (harassment) may act to make systems more spheroidal, and processes like ram-pressure stripping may act to suppress gas accretion (Moore et al. 1998; Abadi et al. 1999; Choi & Yi 2017) which would otherwise spin galaxies up.

It is worth noting that, while the global morphological trend in the Universe is for discs to transform into spheroids, the reverse transformation is also possible (in individual events) through the accretion of gas, as this gas settles into rotational orbits and creates stars that add to the rotational component of the system. Indeed, in very gas-rich major mergers the residual gas may reform a disc, so that the remnant may be discy rather than spheroidal (e.g. Springel & Hernquist 2005; Hau et al. 2008; Kannappan et al. 2009; Font et al. 2011; Aumer et al. 2013; Rodrigues et al. 2017; Sparre & Springel 2017). At high redshift ($z > 2$), cosmological accretion likely plays a dominant role in building up and reforming discs, especially in galaxies fainter than $L^*$ (e.g. Brooks et al. 2009). During these epochs the dominant fuel for star formation and source of angular momentum acquisition in discs are filamentary inflows of cold gas, rather than accretion of shock-heated gas or hierarchical merging (e.g. Murali et al. 2002; Kereš et al. 2005; Brooks et al. 2009; Kimm et al. 2011; Pichon et al. 2011; Stewart et al. 2013; Martin et al. 2016; Welker et al. 2017). Coherent cold flows appear capable of reforming discs up to a critical mass of $10^{10.5}$ M$_\odot$, after which the coherence of the flow is lost and the galaxy morphology is frozen in (Welker et al. 2017).

The orbital parameters of mergers and the spins of accreted satellites may also be an important factor in the morphological evolution of galaxies (e.g Taylor et al. 2018). The alignment or misalignment of both the orbit and the spin of the satellite, relative to the spin of a massive accreting galaxy, may be an important factor in determining the evolution of their angular momentum at later times. The orbits of satellites have been shown to align progressively with the major axis (e.g. Yang et al. 2006) and spin (e.g. Ibata et al. 2013; Welker et al. 2014, 2015) of the more massive galaxy during infall. A preference for prograde mergers may be important for the survival of discs, because in cases where the satellite's orbit is in the same direction as the spin of the more massive merging companion, a merger remnant where disc morphology is preserved may be more probable (e.g. Hopkins et al. 2009b).

Observational studies generally support the predictions of theoretical work. For example, broad morphological change from discs to spheroids has been observed in many studies, across a range in redshift (e.g. Butcher & Oemler 1984; Dressler et al. 1997; Conselice et al. 2014; Huertas-Company et al. 2015). Many spheroids show signatures of violent and sudden morphological change in their stellar populations (e.g. Blake et al. 2004; Bundy et al. 2005; Goto 2005; Kaviraj et al. 2008, 2009, 2011; Kaviraj 2014a; Wild et al. 2016), internal dynamics (e.g. Tacconi et al. 2008; Perret et al. 2012; Cappellari 2016; Rodrigues et al. 2017) and structure (e.g. McIntosh et al. 2008; Conselice & Arnold 2009; Kaviraj et al. 2012a,b; Huertas-Company et al. 2015, 2016), indicating a major merger in their recent history. However, recent work has also demonstrated that many spheroids (especially at $z \sim 2$) appear to be forming without recourse to major mergers, indirectly supporting the potentially important role of minor mergers in driving morphological transformation (e.g. Bundy et al. 2007; Pracy et al. 2009; Kaviraj et al. 2013; Haines et al. 2015; Lofthouse et al. 2017).

Nevertheless, while today's surveys are able to provide datasets of sufficient quality that is possible for galaxy populations across a large range in redshift to be compared morphologically, an empirical determination of the role that mergers and other processes may play in the morphological evolution of galaxies remains difficult. For example, given the limited depth and/or survey area of past surveys, samples of mergers are typically small (e.g. Darg et al. 2010a,b). And since the surface brightness of tidal features induced by mergers decreases with the mass ratio of the merger (e.g. Peirani et al. 2010), most surveys are too shallow to detect the signatures of low mass ratio mergers (see e.g. Kaviraj 2013; Kaviraj 2014b). Furthermore, disturbed morphologies may result naturally from internal processes, especially in the early Universe (e.g. Bournaud et al. 2008; Agertz et al. 2009; Förster Schreiber et al. 2011; Cibinel et al. 2015; Hoyos et al. 2016), making it difficult to accurately separate merger remnants from the non-interacting population. Thus, even as we enter an era of deep-wide observational surveys (e.g. DES (Dark Energy Survey Collaboration et al. 2016), EUCLID (Laureijs et al. 2011), LSST (Tyson 2002; Robertson et al. 2017) and JWST (Gardner et al. 2006)), a purely empirical study of the processes that contribute to the morphological evolution of galaxies remains a challenge.

While theoretical studies offer a better avenue for exploring morphological transformation, many theoretical explorations of this issue have focussed on isolated and idealised simulations of galaxy mergers (e.g. Barnes 1988; Hernquist 1992; Bois et al. 2011). However, such simulations lack a realistic context, and so exclude the effects of environment and gas accretion from the cosmic web. Additionally, since the parameter space explored by these studies is small and is not informed by a cosmological model, it is not possible to make statistical statements about the importance of mergers and other processes to morphological transformation globally. While 'zoom-in' studies from cosmological simulations (e.g. Sales et al. 2012; Wuyts et al. 2014; Fiacconi et al. 2015; Sparre & Springel 2016) do offer a way of placing merging systems into a realistic environment, without requiring significant increases in computing power, both approaches are generally limited by small sample sizes and restricted parameter spaces. Cosmological volumes are essential for a statistical study of morphological transformation.

In the recent literature, semi-analytical models (e.g. Kauffmann et al. 1993; Somerville et al. 2001; Menci et al. 2002; Hatton et al. 2003; Lu et al. 2011) have played an important role in exploring galaxy evolution, using large, statistically-significant samples. While these models have been able to reproduce broad trends in galaxy formation, including the evolution of morphology, stellar mass and gas content (e.g. Somerville & Primack 1999; Cole et al. 2000; Benson et al. 2003; Bower et al. 2006; Croton et al. 2006; Khochfar et al. 2011; Lamastra et al. 2013; Tonini et al. 2016), they are essentially phenomenological and lack realistic baryonic physics, relying instead on simple numerical recipes for sub-galaxy-scale processes, including morphological transformation. However, recent advances in computing power mean that it has now become possible to simulate the resolved baryonic physics (e.g. gas content and stellar populations) of individual galaxies within cosmological volumes. Modern cosmological, hydrodynamical simulations (e.g. Dubois et al. 2014; Vogelsberger et al. 2014; Khandai et al. 2015; Schaye et al. 2015; Taylor & Kobayashi 2016; Dubois et al. 2016; Kaviraj et al. 2017) are typically capable of resolving baryonic physics on kpc scales, allowing for the detailed study of small-scale processes within large populations of galaxies. Such simulations offer an unprecedented route to understanding the relative role of different mechanisms in driving the evolution of the morphological mix of the Universe (e.g. Welker et al. 2017; Rodriguez-Gomez et al. 2017; Clauwens et al. 2017).

In this paper, we use Horizon-AGN (Dubois et al. 2014; Kaviraj et al. 2017), a cosmological hydrodynamical simulation, to investigate key open questions in our understanding of the evolution of the morphological mix of the Universe: what is the magnitude of morphological change imparted by major and minor mergers as a function of redshift and stellar mass? what is the impact of gas fraction on these morphological changes? are the properties of the remnants dependent on the orbital configurations (e.g. prograde vs retrograde) of mergers? what fraction of the total morphological change over cosmic time is attributable to major and minor mergers and other processes?

The structure of this paper is as follows. In Section 2 we present an overview of Horizon-AGN, outlining the treatment of baryonic physics and black holes, the identification of galaxies and mergers, and the definition of morphology used in this study. In Section 3 we explore the effect that individual mergers have in driving changes in morphology as a function of redshift, merger mass ratio, gas fraction and orbital configuration. In Section 4 we study the average merger histories of discs and spheroids, quantify the cumulative effect of major and minor mergers over cosmic time and outline the role of environment in producing morphological change in regions of high density (e.g. clusters). We summarise our findings in Section 5.

## 2 THE SIMULATION

Horizon-AGN is a cosmological-volume hydrodynamical simulation (Dubois et al. 2014), based on RAMSES (Teyssier 2002), an adaptive mesh refinement (AMR) Eulerian hydrodynamics code. It simulates a 100 $h^{-1}$ coMpc length box, using initial conditions from a *WMAP7* ΛCDM cosmology (Komatsu et al. 2011). The simulation contains $1024^3$ dark matter particles, with a mass resolution of $8 \times 10^7$ M$_\odot$. An initially uniform $1024^3$ cell gas grid is refined, according to a quasi Lagrangian criterion (when 8 times the initial total matter resolution is reached in a cell) and the refinement can continue until a minimum cell size of 1 kpc in proper units is reached.

As shown in Kaviraj et al. (2017) and Kaviraj et al. (2015), Horizon-AGN produces good agreement to key observables that trace the aggregate evolution of galaxies across cosmic time e.g. stellar mass and luminosity functions, rest-frame UV-optical-near infrared colours, the star formation main sequence, galaxy merger histories and the cosmic star formation history. It also reproduces the demographics of BHs, including BH luminosity and mass functions, the BH mass density as a function of redshift, and correlations between BH and galaxy mass in the local Universe (Volonteri et al. 2016). Finally, Horizon-AGN reproduces the morphological mix of the local Universe, with predicted galaxy morphologies in good agreement with observed morphological fractions for intermediate and high mass galaxies (Dubois et al. 2016; Martin et al. 2018a).

In the following sections, we briefly describe aspects of the simulation that are particularly relevant to this study: the treatment of baryonic matter (gas and stars), the identification of galaxies and mergers, the measurement of galaxy morphology and the treatment of BHs and BH feedback on ambient gas.

### 2.1 Baryons

Gas cooling proceeds via H, He and metals (Sutherland & Dopita 1993) down to a temperature of $10^4$ K and a uniform UV background is switched on at $z = 10$, following Haardt & Madau (1996). Star formation is implemented via a standard 2 per cent efficiency (e.g. Kennicutt 1998), when the density of hydrogen gas reaches 0.1 H cm$^{-3}$. The stellar-mass resolution in the simulation is $\sim 2 \times 10^6$ M$_\odot$.

Continuous stellar feedback is employed, including momentum, mechanical energy and metals from stellar winds and Type II and Type Ia supernovae (SNe). Energetic feedback from stellar winds and Type II SNe is applied via STARBURST99 (Leitherer et al. 1999, 2010), implemented via the Padova model (Girardi et al. 2000) with thermally pulsating asymptotic branch stars (Vassiliadis & Wood 1993). The kinetic energy of stellar winds is calculated via the 'Evolution' model of Leitherer et al. (1992). The implementation of Type Ia SNe follows Matteucci & Greggio (1986) and assumes a binary fraction of 5% (Matteucci & Recchi 2001), with chemical yields taken from the W7 model of Nomoto et al. (2007). Stellar feedback is modelled as a heat source after 50 Myrs. This is because after 50 Myrs the bulk of the energy is liberated via Type Ia SNe that have time delays of several hundred Myrs to a few Gyrs (e.g. Maoz et al. 2012). These systems do not suffer large radiative losses, as stars disrupt or move away from their dense birth clouds after around a few tens of Myrs (see e.g. Blitz & Shu 1980; Hartmann et al. 2001).

### 2.2 Identification of galaxies and mergers

The ADAPTAHOP structure finder (Aubert et al. 2004; Tweed et al. 2009), applied to the distribution of star particles, is used to identify galaxies. The selection of structures requires that the local density exceeds 178 times the average matter density. The local density is calculated using the 20 nearest particles. A minimum number of 50 particles is required to identify a structure, which imposes a minimum galaxy stellar mass of $2 \times 10^8$ M$_\odot$. We produce merger trees for each individual galaxy and track their progenitors to $z = 3$. The average timestep in the merger histories is $\sim 130$ Myr. A major merger is defined as a merger where the mass ratio of the merging progenitors is greater than or equal to $1 : 4$. A minor merger is defined as a merger where the mass ratio of the merging progenitors is between $1 : 4$ and $1 : 10$.

The choice of a threshold mass ratio for minor mergers of $1 : 10$ is not arbitrary, but driven by previous work which indicates that this is typically a threshold below which the impact of mergers generally becomes negligible. For example, below this threshold, star formation and black-hole accretion rates are not detectably enhanced in mergers (Martin et al. 2017; Martin et al. 2018b). In the Appendix, we demonstrate this point by quantifying the effect of varying the minimum mass ratio down to $1 : 40$ for our results in Sections 3 and 4, and showing that, in mergers with mass ratios below 1 : 10, there is negligible morphological change, compared to galaxies that are not merging.

The requirement of 50 particles for the definition of a galaxy, imposes a limit on the minimum merger mass ratio that is detectable for a galaxy of a given mass at a given redshift. Since galaxies contain less stellar mass at higher redshift, we detect a smaller proportion of mergers at earlier times. Figure 1 presents detectability limits for mergers of various mass ratios in the merger histories of galaxies, as a function of the stellar mass of galaxies at $z = 0$. For each galaxy, we calculate the mass of its main progenitor at a redshift of interest. This then determines the mass ratio limit of detectable mergers for the galaxy in question.

In Figure 1, we show the fraction of galaxies of a given stellar-mass at $z = 0$ that have progenitors that are massive enough for mergers of various mass ratios to be detectable at different redshifts. For example, for galaxies that have a mass of $10^{9.5}$ M$_\odot$ at the present day, 96, 72 and 37 per cent of their progenitors are massive enough for a merger with a 1 : 4 mass ratio to be detectable at $z = 1$, $z = 2$ and $z = 3$ respectively. For mergers with mass ratios of 1 : 10, the corresponding values are 84, 47 and 20 per cent at the same redshifts. For galaxies with stellar masses above $10^{11}$ M$_\odot$, the merger history is at least 85 per cent complete for mass ratios greater than 1 : 10, up to $z = 3$. We note that, while mergers will not be detectable for a large proportion of very low-mass galaxies, the merger rate is expected to fall with decreasing stellar mass (e.g. Stewart et al. 2008; Rodriguez-Gomez et al. 2015), so that the importance of mergers is lower in the regime where the sample is most incomplete.

## 2.3 Galaxy morphology

Following Martin et al. (2018a), we estimate morphology using a galaxy's stellar kinematics. We use $V/\sigma$, the ratio of the mean rotational velocity ($V$) and the mean velocity dispersion ($\sigma$), measured using the entire star particle distribution of the galaxy. Higher values of $V/\sigma$ correspond to systems that are more rotationally-supported i.e. those that have more late-type (disc-like) morphologies. $V/\sigma$ is calculated by rotating the coordinate system so that the $z$-axis is oriented along the stellar angular momentum vector. $V$ is then defined as the mean tangential velocity component in cylindrical co-ordinates, $V_\theta$. The velocity dispersion ($\sigma$) is computed using the standard deviations of the radial, tangential and vertical star particle velocities, $\sigma_r, \sigma_\theta$ and $\sigma_z$, summed in quadrature. $V/\sigma$ is defined as:

$$V/\sigma = \frac{\sqrt{3}\bar{V}_\theta}{\sqrt{\sigma_r^2 + \sigma_\theta^2 + \sigma_z^2}}. \tag{1}$$

As in Martin et al. (2018a), we use 'spheroid' and 'disc' to refer to galaxies that are dominated by their dispersional and rotational velocities respectively. Following Martin et al. (2018a), we choose a $V/\sigma$ threshold value of 0.55, which best reproduces the observed spheroid and disc fractions of the Universe at low redshift (Conselice 2006). In other words, galaxies with $V/\sigma$ values above 0.55 are considered to be discs, while those with values below this threshold are spheroids.

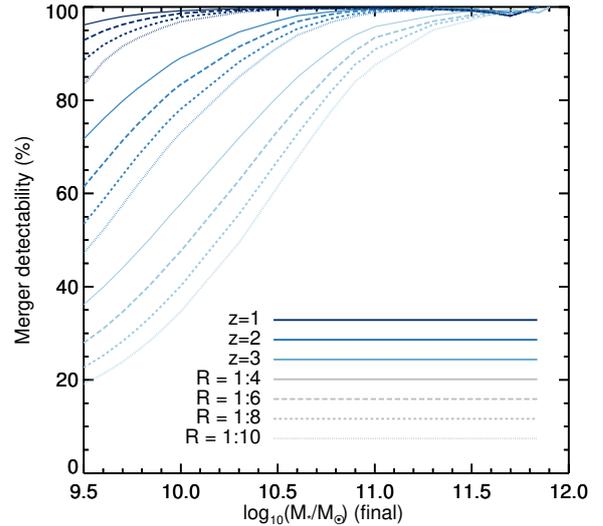

**Figure 1.** The proportion of galaxies for which mergers of different mass ratios are detectable at various redshifts (see legend), as a function of the stellar mass of the galaxy at $z = 0$. $R$ is the stellar mass ratio of the merger. For example, for galaxies that have a mass of $10^{9.5}$ M$_\odot$ at the present day, 96, 72 and 37 per cent of these systems have progenitors that are massive enough for a merger with a 1 : 4 mass ratio to be detectable at $z = 1$, $z = 2$ and $z = 3$ respectively. For mergers with mass ratios of 1 : 10, the corresponding values are 84, 47 and 20 per cent at the same redshifts.

## 2.4 Treatment of black holes and black-hole feedback

BH are seeded as 'sink' particles with an initial mass of $10^5$ M$_\odot$ until $z = 1.5$, wherever the local gas density exceeds $\rho > \rho_0$ and the stellar velocity dispersion exceeds 100 km s$^{-1}$, where $\rho_0 = 1.67 \times 10^{-25}$ g cm$^{-3}$ and corresponds to 0.1 H cm$^{-3}$ (the minimum density threshold required for star formation). To prevent multiple BHs from forming within the same galaxy, BHs cannot form if there is another BH within 50 kpc. Each BH grows through gas accretion, or coalescence with another black hole (Dubois et al. 2014, 2016). An Eddington-limited Bondi-Hoyle-Lyttleton rate is used to model BH accretion:

$$\dot{M}_{BH} = \frac{4\pi\alpha G^2 M_{BH}^2 \bar{\rho}}{(\bar{c}_s^2 + \bar{u}^2)^{3/2}}, \tag{2}$$

where $M_{BH}$ is the mass of the BH, $\bar{\rho}$ is the mass-weighted average gas density, $\bar{c}_s$ is the mass-weighted average sound speed, $\bar{u}$ is the mass-weighted average gas velocity relative to the BH and $\alpha$ is a dimensionless boost factor accounting for the inability of the simulation to capture the cold (high-density) inter-stellar medium (e.g. Booth & Schaye 2009).

BHs impart feedback on gas via two different channels that depend on the ratio of the gas accretion rate and the Eddington luminosity, $\chi = \dot{M}_{BH}/\dot{M}_{Edd}$. For Eddington ratios $\chi > 0.01$ (that represent high accretion rates), BH feedback operates via a 'quasar' mode, with 1.5 per cent of the accretion energy injected as thermal energy into the gas isotropically. For Eddington ratios $\chi < 0.01$ (which represent low accretion rates), BH feedback is modelled as a 'radio' mode, where bipolar outflows are implemented with jet velocities of $10^4$ km s$^{-1}$. The efficiency of the quasar mode is chosen to reproduce the local observed $M_{BH} - M_\star$ and $M_{BH} - \sigma_\star$ re-lations and the local cosmic black-hole mass density (Dubois et al. 2012).

BH feedback principally quenches external gas accretion (e.g. Dubois et al. 2014, 2016), but does not couple to star particles. Gas accretion typically acts to increase $V$, since new stars forming from the gas inherit its angular momentum and add to the rotational component of the galaxy. Thus, the effect of BH feedback is essentially to prevent the value of $V$ from increasing and, therefore, to also lock in the value of $V/\sigma$, since it doesn't alter $\sigma$, the stellar velocity dispersion. BH feedback therefore plays an important, if indirect role, in the morphological evolution of galaxies. It is necessary for reproducing the observed morphological diversity of the present-day Universe (e.g. Dubois et al. 2016), as it can *maintain* the morphology of a system, in between events like mergers that alter it (e.g. Dubois et al. 2016; Pontzen et al. 2017). It is worth noting, however, that BH feedback cannot, by itself, create spheroidal systems. Indeed in Horizon-AGN, we do not find significant differences in the aggregate BH accretion rates in spheroid and disc progenitors, again indicating that BH feedback alone is not capable of producing spheroidal morphologies.

## 3 THE EFFECT OF INDIVIDUAL MERGERS

We begin by investigating how key properties of the progenitors affect the morphology of the merger remnant. The large volume of the Horizon-AGN simulation allows us to explore a realistic Λ CDM parameter space of mergers, across a broad range of properties, including redshift, stellar mass, merger mass ratio, gas fraction and orbital configuration. In this section, we first describe our method for calculating the change in morphology due to a merger, study the effect of major and minor mergers on the morphology of galaxies as a function of stellar mass and redshift (Section 3.1), explore the effect of gas fraction on the morphology of merger remnants (Section 3.2) and probe how orbital configurations influence the properties of merger remnants (Section 3.3).

In order to quantify the morphological change in a merging system, we measure the change in $V/\sigma$ of the main (i.e. the more massive) progenitor in a 2 Gyr window, centred around the time that the two galaxies coalesce (i.e. when both galaxies are identified as being part of the same structure). As we elaborate below, the size of the window is chosen to ensure that we measure the morphology of the main companion before it is affected by gravitational torques in the merger, and to allow time for the merger remnant to relax, at least in its inner regions (low-surface-brightness features in galaxy outskirts, such as shells and loops, can last for many dynamical timescales, e.g. Mihos (2000); Kaviraj (2014b)), but make up a negligible proportion of the galaxy's stellar mass). Note that, since we consider the collection of individual merger events in this section, incompleteness does not affect our analysis, on the assumption that the statistical properties of mergers that are not visible are similar to those that are observed. Incompleteness is a larger issue when studying the cumulative impact of mergers over cosmic time, and we return to this point in the next section.

For each galaxy merger, we measure the morphological change, $\Delta$morph, defined as the fractional change in the $V/\sigma$ of the main progenitor over the course of the merger. We assume a timescale of 2 Gyrs, measuring the change in $V/\sigma$ between $t = -1$ Gyr and $t = +1$ Gyr relative to coalescence:

$$\Delta\text{morph} = \frac{V/\sigma_{t=1\text{ Gyr}} - V/\sigma_{t=-1\text{ Gyr}}}{V/\sigma_{t=-1\text{ Gyr}}},\qquad(3)$$

where $\Delta t = t_{1\text{ Gyr}} - t_{-1\text{ Gyr}}$, and is approximately equal to 2 Gyrs, with the exact value depending on the coarseness of the merger-tree timesteps. The choice of a 2 Gyr timescale is driven by the fact that, for the merger mass ratios we will consider in this study ($>1$:10), the merger process is typically complete over this timescale (e.g. Jiang et al. 2008; Kaviraj et al. 2011). We note that an important issue when selecting a timescale is to use a value that encompasses the merger event completely. In particular, choosing timescales that are too short will lead to spurious results, because merger remnants may not have relaxed at the point at which they are observed. We explore our choice of timescale in more detail in the Appendix and show that choosing a slightly longer 3 Gyr or even a 1 Gyr timescale does not alter the conclusions of this paper, although, as we discuss below, 1 Gyr may be too short for mergers closer to the lower end ($\sim 1$ : 10) of our mass ratio range.

Finally, galaxies are considered to be not merging if they have not undergone either a major or minor merger within the last Gyr, and will not undergo such a merger in the next Gyr. All galaxy properties, such as $M_\star$, $m_{gas}$ etc., are calculated at the initial ($t = -1$ Gyr) snapshot. We also calculate $\Delta$morph for galaxies that are not undergoing mergers in the same way as for the merging galaxies, again using a timescale of 2 Gyr.

### 3.1 Morphological change induced by major and minor mergers as a function of stellar mass and redshift

In Figure 2, we investigate $\Delta$morph as a function of the stellar mass and redshift of the main (more massive) progenitor. Galaxies undergoing major and minor mergers are indicated using the solid and dashed lines respectively, while galaxies that are not undergoing any mergers are indicated using the dotted lines. Recall that a major merger is defined as a merger where the mass ratio of the merging progenitors is greater than or equal to 1 : 4, while a minor merger is defined as one where the mass ratio of the merging progenitors is between 1 : 4 and 1 : 10. We further separate the main progenitors into spheroids (left-hand column) and discs (right-hand column). Positive values of $\Delta$morph indicate that the merger remnant has spun up (i.e. become more rotationally-supported or discy), while negative values of $\Delta$morph indicate that the remnant has spun down (i.e. become more dispersion-supported or spheroidal).

Mergers in which the main progenitor is a disc galaxy almost exclusively spin down, and result in remnants with lower $V/\sigma$, i.e. systems that are more spheroidal. Spinning up as a result of such mergers is rare and happens only in ~5 per cent of cases for main progenitor masses of $10^{10.5} < M_\star/\text{M}_\odot < 10^{11}$. The values of $\Delta$morph in Figure 2 indicate that major mergers where the main progenitor is a disc galaxy produce larger morphological changes than minor mergers. In the nearby Universe, individual major and minor events with main progenitor masses of $10^{10.5} < M_\star/\text{M}_\odot < 10^{11}$ reduce $V/\sigma$ by around ~28 per cent and ~13 per cent respectively over the course of the merger. The corresponding values at $z \sim 1$ and $z \sim 2$ are 44/27 per cent, 57/39 per cent respectively. Typically, the magnitude of the morphological change induced by major mergers is around a factor of 2 greater than that in minor mergers.

While the effect of individual mergers is largely insensitive to the stellar mass of the main progenitor, it is dependent on the dispersional component of the stellar velocity distribution in a galaxy. As galaxies grow larger bulge components towards the present day, more of a merger's ability to induce morphological change is removed, as larger proportions of stellar mass have already been removed from circular orbits and re-arranged into random orbits. This leads to the gradual decrease in Δmorph in discs towards lower redshift.

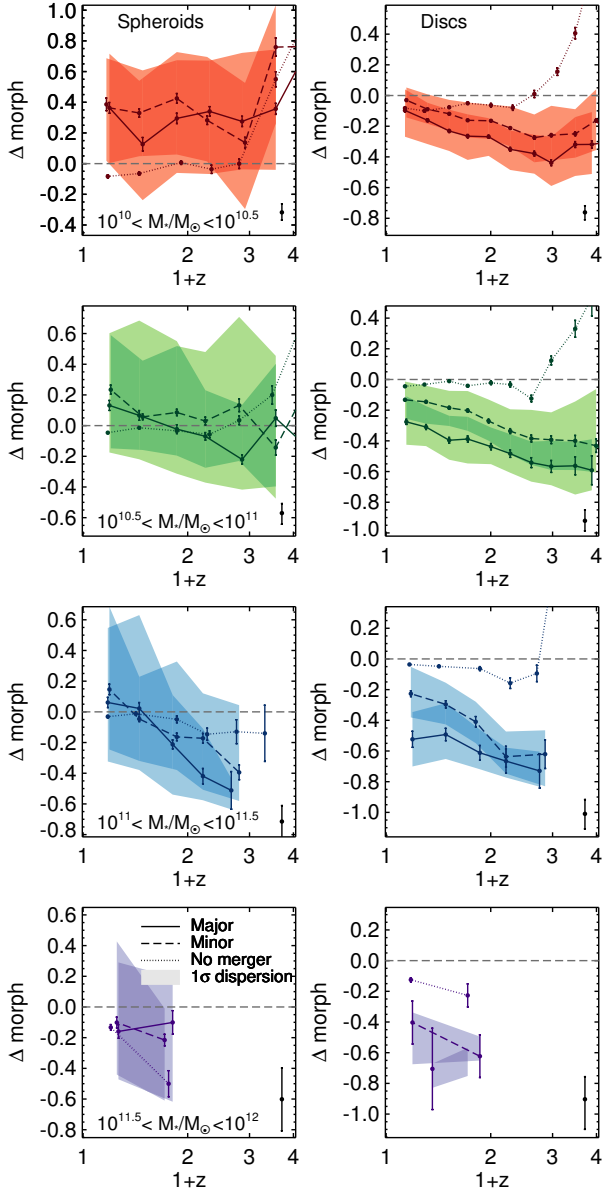

**Figure 2.** Median Δmorph as a function of redshift and stellar mass. Positive values of Δmorph indicate that the merger remnant has spun up (i.e. become more rotationally-supported), while negative values of Δmorph indicate that the remnant has spun down (i.e. become more dispersion-supported). The left-hand column indicates mergers where the main (i.e. more massive) progenitor is a spheroid, while the right-hand column shows mergers where the main progenitor is a disc. Error bars indicate the standard error on the median and filled regions indicate ±1σ dispersions. For non-merging galaxies, typical dispersions are indicated by a black error bar. The solid and dashed dark lines indicate the median Δmorph for major and minor mergers respectively, while the dotted line indicates Δmorph for non-merging galaxies. Galaxies with $M_\star > 10^{11.5} M_\odot$ only begin to appear in the simulation after $z \sim 2$, so there are no datapoints for higher redshifts.

The impact of mergers on spheroids is qualitatively different. Unlike discs, which are efficiently destroyed by major and minor mergers, there is little preference for spin down over spin up in spheroids. As indicated by the 1σ dispersion regions, mergers where the main progenitor is a spheroid produce a greater range of outcomes, and can produce both positive and negative values for Δmorph. In high redshift ($z > 1$) mergers that involve massive spheroidal galaxies ($M_\star/M_\odot > 10^{11}$), mergers tend, on average, to spin remnants down. The magnitude of this morphological change is similar to what is observed in mergers where the main progenitor is a disc, although the scatter in the range of Δmorph values is higher, so that it becomes more likely that some mergers spin remnants up (this happens in approximately 30 per cent of major mergers and 40 per cent of minor mergers).

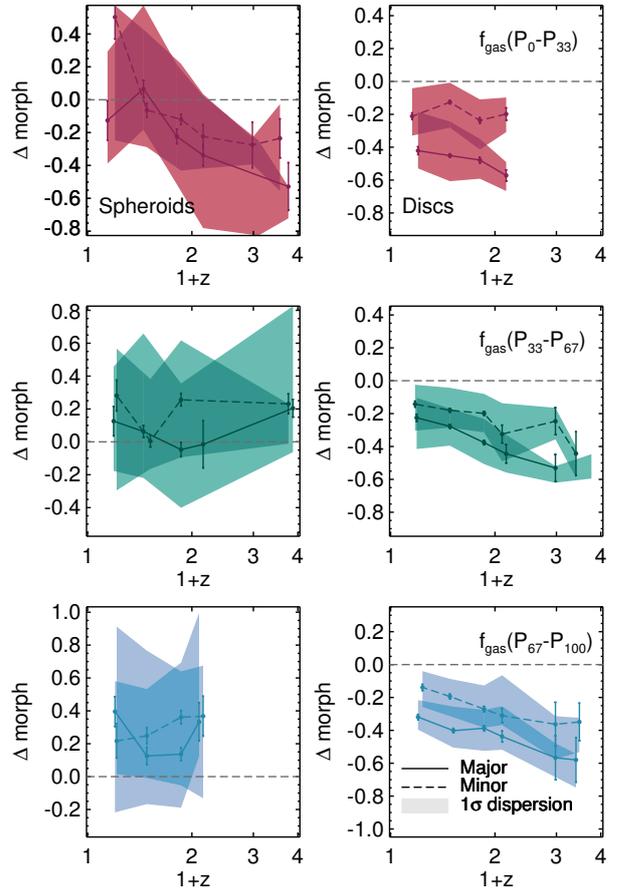

**Figure 3.** Median Δmorph as a function of redshift and split by the gas fraction percentile they inhabit for galaxies in the mass range $10^{10.5} < M_\star/M_\odot < 10^{11}$. The 33rd and 67th percentiles are used to create the percentile ranges, and are typically 0.85 times and 1.15 times the average values indicated by the green lines in the top panel of Figure 4. The left-hand column indicates mergers where the main (i.e. more massive) progenitor is a spheroid, while the right-hand column shows mergers where the main progenitor is a disc. Error bars indicate the standard error on the median, while filled regions indicate the ±1σ dispersions. The solid and dashed dark lines indicate the median Δmorph for major and minor mergers respectively. Note that the behaviour is similar regardless of the mass of the main progenitor.

In low redshift ($z < 1$) mergers that involve such massive spheroidal galaxies, the average value of Δmorph is close to zero i.e. mergers do not, on average, produce strong morphological changes, although, as the 1σ dispersion regions indicate, both spinning up and spinning down is possible from such events. The impact of mergers that involve less massive spheroidal main progenitors is somewhat different. While for intermediate stellar masses ($10^{10.5} < M_\star/M_\odot < 10^{11}$) the average values of Δmorph remain close to zero across all redshifts, at the low-mass end ($10^{10} < M_\star/\mathrm{M}_\odot < 10^{10.5}$) remnants of both major and minor mergers tend to spin remnants up.

For non-merging galaxies, similar trends are observed in mergers where the main progenitors are spheroids or discs, largely regardless of stellar mass. $\Delta$morph is large and positive at high redshift ($z > 2$), indicating intense cosmological gas accretion that imparts angular momentum to the galaxy and spins it up (e.g. Brooks et al. 2009; Pichon et al. 2011; Stewart et al. 2013). At lower redshifts, $\Delta$morph in non-merging galaxies is close to zero (indicating no morphological change), but is typically slightly negative ($\Delta$morph $\gtrsim -0.08$) within errors, possibly indicating some morphological impact from very low mass ratio ($< 1 : 10$) interactions.

## 3.2 The effect of gas fraction

In Figure 3, we study the value of $\Delta$morph as a function of the redshift and gas fraction of the merging system. We again separate our analysis into main progenitors that are spheroids (left-hand column) and discs (right-hand column). The gas fraction ($f_{gas}$) is defined as the combined cold gas fraction of the two merging companions:

$$f_{gas} = \frac{m_{gas,main} + m_{gas,sat}}{m_{gas,main} + m_{\star,main} + m_{gas,sat} + m_{\star,sat}},\qquad(4)$$

where $m_{gas,main}$ is the mass of cold gas within 2 $R_{\mathrm{eff}}$ of the main (more massive) companion 1 Gyr prior to coalescence, and $m_{gas,sat}$ is the corresponding value for the lower mass companion (i.e. the satellite). Similarly, $m_{\star,main}$ is the stellar mass within 2 $R_{\mathrm{eff}}$ of the more massive companion and $m_{\star,sat}$ is the corresponding value for the smaller companion. For this analysis, we restrict our sample to a narrow range in stellar mass, $10^{10.5} < M_\star/\mathrm{M}_\odot < 10^{11}$, since the mean gas fraction evolves with galaxy stellar mass. However, the behaviour we find here is similar regardless of the stellar mass range considered. We split our sample by gas fraction percentile after splitting the sample by redshift and the morphology of the main progenitor. The top and bottom panels represent the extreme objects in terms of gas fraction for a given redshift. The 33rd and 67th percentiles are used to create the percentile ranges, and are typically 0.85 times and 1.15 times the average values indicated by the green lines in the top panel of Figure 4.

Mergers where the main progenitor is a disc galaxy typically produce remnants with negative $\Delta$morph i.e. typically spin systems down, largely regardless of the gas fraction. However, the behaviour in spheroids is different. While mergers with low gas fractions are most likely to spin remnants down (apart from in the nearby Universe, where spin up or down appear equally likely), mergers that involve high gas fractions produce significant spin up, particularly in the high-redshift Universe, in line with the results of recent observational and theoretical work (e.g. Robertson et al. 2006; Athanassoula et al. 2016; Rodriguez-Gomez et al. 2017; Font et al. 2017).

Figure 4, which shows the evolution of the average galaxy gas fraction over cosmic time, helps explain the mass dependence of $\Delta$morph for discs and spheroids seen in Figure 2. As this figure indicates, average gas fractions evolve as a function of mass and redshift, with lower mass galaxies exhibiting higher values, consistent with the results of recent observational studies (e.g. Geach et al. 2011; Tacconi et al. 2013). Lower mass galaxies tend to be more gas-rich (at any epoch). Since high gas fractions tend to produce remnants with more positive $\Delta$morph values (Figure 3),

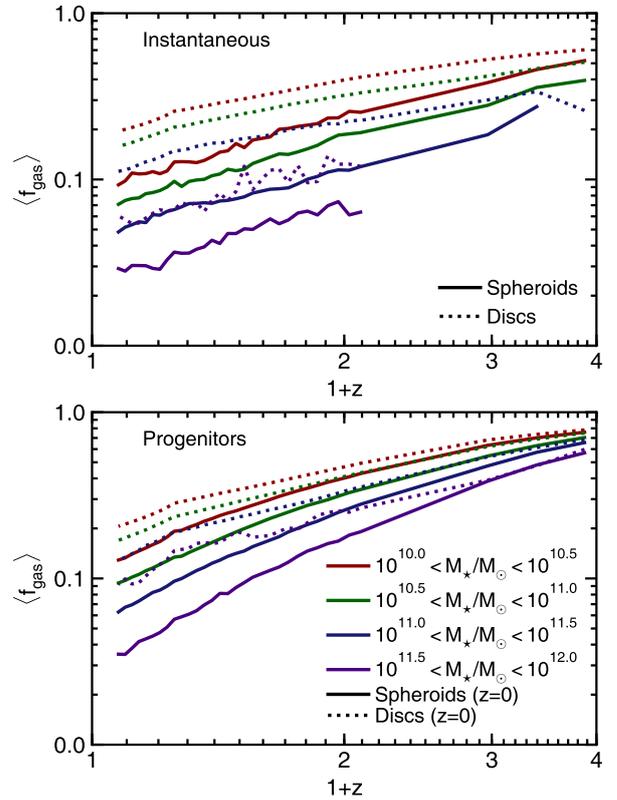

**Figure 4.** The redshift evolution of the mean cold gas fraction ($f_{gas}$) in spheroidal and disc galaxies in bins of stellar mass. **Top:** Mean value of $f_{gas}$ in spheroids (solid lines) and discs (dotted lines) as a function of redshift, shown in bins of stellar mass. **Bottom:** Mean value of $f_{gas}$ in the main (i.e. the most massive) progenitors of today's spheroids (solid lines) and discs (dotted lines), shown as a function of the stellar mass of the final spheroid or disc at $z = 0$.

i.e. milder morphological transformation, mergers involving lower mass galaxies are more gas rich (e.g. Kaviraj et al. 2009; Struve et al. 2010), which explains the trend in $\Delta$morph becoming more positive for both discs and spheroids at lower stellar masses.

## 3.3 The effect of orbital configuration

We proceed by investigating how $\Delta$morph varies as a function of the orbital configuration of the merging system, at the point where the satellite enters the virial radius of the main (more massive) progenitor. We consider both the angular momentum of the satellite galaxy, $\boldsymbol{L}_{sat}$, measured using its star particle distribution, and the angular momentum of the orbit of the satellite relative to the main progenitor, $\boldsymbol{L}_{orb} = M_{sat}(\boldsymbol{r} \times \boldsymbol{v})$, where $\boldsymbol{r}$ and $\boldsymbol{v}$ are the position and velocity of the satellite relative to that of the main progenitor. The meaning of the angular momentum vectors is illustrated in Figure 5. We use these quantities to define the total angular momentum that is imparted by the merging satellite to the main progenitor, relative to its spin:

$$L_{external} = |\boldsymbol{L}_{sat}|\cos(\theta_{L_{main},L_{sat}}) + |\boldsymbol{L}_{orb}|\cos(\theta_{L_{main},L_{orb}})\qquad(5)$$

where $\theta_{L_{main},L_{sat}}$ is the angle between the angular momentum vector of the main progenitor and the satellite, so that an angle below $\pi/2$ denotes two co-rotating galaxies (i.e. a 'prograde' merger), while angles above $\pi/2$ denote counter-rotating galaxies (i.e. a 'retrograde' merger). $\theta_{L_{main}, L_{orb}}$ is the angle between the angular momentum vector of the main progenitor and the angular momentum vector of the satellite's orbit. Thus, an angle that is close to zero denotes a merger where the orbit of the merging satellite is co-planar with the disc of the main progenitor and in the same direction as its rotation, while a value of $\pi$ denotes a merger that impacts the disc of the main progenitor in the opposite direction to its rotation, most efficiently removing angular momentum from the system. Note that our definition of prograde or retrograde includes both the angular momentum of the satellite's orbit as well as its spin relative to the spin of the main progenitor.

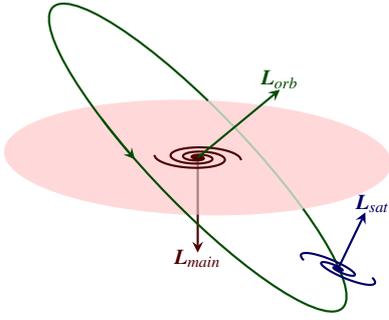

**Figure 5.** Sketch illustrating the vectors used to describe the orbital configuration of merging systems. The main progenitor refers to the more massive merging companion. The angles between vectors are defined in the standard way as the angles between two vectors in their common plane.

Figure 6 shows the average $\Delta$morph in prograde and retrograde mergers where the main progenitor is a disc galaxy, as a function of redshift. Major mergers efficiently destroy discs in almost all cases, but are most effective when the angular momentum of the spin of the satellite and its orbit is counter to the rotation of the main progenitor's disc (i.e. $L_{external} < 0$). The difference is fairly modest, especially at high redshift, where major mergers produce a mean $\Delta$morph of 0.52 and 0.50 for retrograde and prograde mergers respectively. The difference is more significant at low redshifts (mean $\Delta$morph of 0.35 and 0.20 for retrograde and prograde mergers) when average gas fractions are low (see Figure 4) and mergers are less effective at rebuilding discs. Low-redshift minor mergers are almost equally likely to produce positive or negative changes to $\Delta$morph when the merger is prograde. In the case of retrograde minor mergers, spin down of discs remains significant in almost all cases.

Figure 7 shows the fraction of prograde events in mergers that involve progenitors of today's spheroids (solid line) and today's discs (dotted line). The prograde fractions are shown as a function of the stellar mass of the spheroid or disc at $z = 0$. The progenitors of today's discs have undergone more prograde mergers than the progenitors of spheroids in all but the highest stellar mass bin, where the prograde fractions are relatively similar within errors (note that there are very few discs with such high stellar masses, which leads to a large error bar in the prograde fraction). The tendency of disc progenitors to have had more prograde mergers over cosmic time is therefore likely to be a contributing factor to the continued survival of these discs.

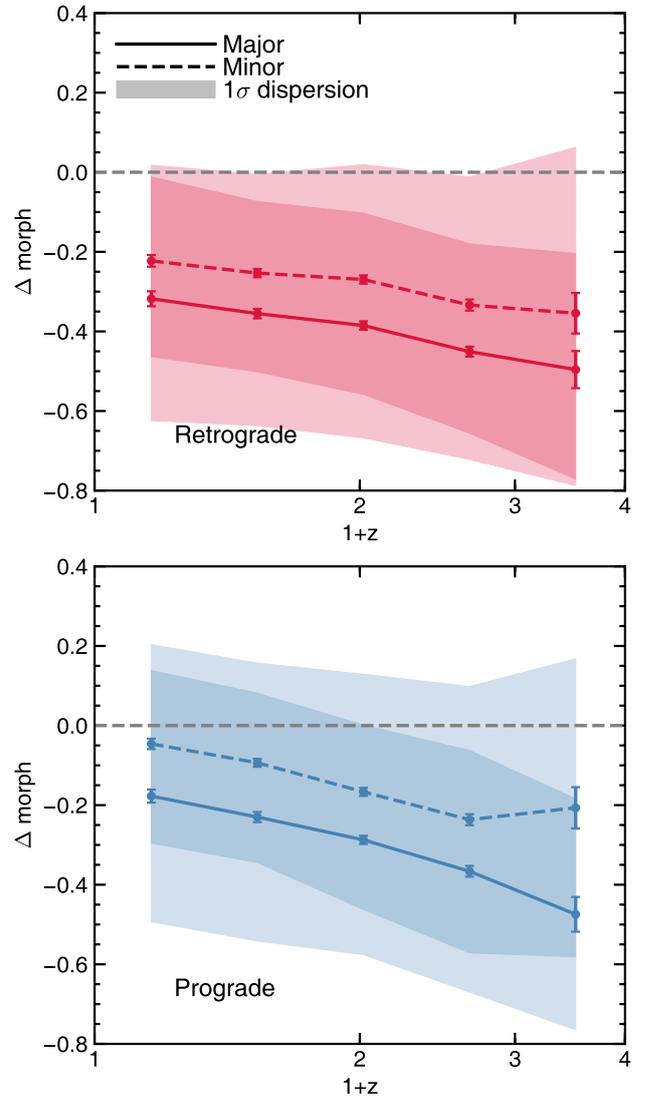

**Figure 6.** Median $\Delta$morph for discs as a function of redshift for prograde ($L_{external} > 0$) and retrograde ($L_{external} < 0$) mergers. The left-hand column indicates retrograde mergers, where the angular momentum is imparted counter to the spin of the main (more massive) merging progenitor and the right-hand column indicates prograde mergers where the angular momentum is imparted in the same direction as the spin of the main progenitor. Error bars indicate the standard error on the median and solid and dashed lines indicate the median $\Delta$morph for major mergers and minor mergers respectively.

## 3.4 Average $\Delta$morph in the progenitors of today's discs and spheroids

While the previous sections have explored the impact of mass ratios, gas fractions and orbital configurations, we complete our analysis of the effect of individual mergers by taking an average, aggregate view on the morphological transformation that takes place in mergers undergone by the *progenitors* of present-day discs and spheroids.

Figure 8 shows the mean $\Delta$morph in events that constitute the merger history of massive ($M_\star/M_\odot > 10^{10.5}$) systems at the present day. In other words, while Figure 2 considered the morphology of the main progenitors at the time of the merger itself, Figure 8 considers the set of events that make up the merger history of galaxies that are spheroids and discs at the present day. Typically, we find that events in the merger history of today's spheroids spin remnants down. For most of cosmic time, major mergers produce $\Delta$morph values of 0.5, while minor mergers produce values of 0.3. On the other hand, events in the merger history of today's discs produce much smaller morphological changes, with $\Delta$morph values of less than 0.2 and 0.1 for major and minor mergers respectively. The discrepancy between the average outcomes of mergers between the progenitors of spheroids and discs shows that the morphological evolution of galaxies must be shaped to some extent by the properties of the merging galaxies themselves, particularly the cold gas fraction and the direction of the angular momentum of the spin and orbit of the merging satellite with respect to the spin of the more massive progenitor.

**Table 1.** Mean (median) properties of spheroids (white rows) and discs (shaded rows) at $z = 0$. ± indicates the $1\sigma$ dispersion. (i) Number of galaxies in mass bin, (ii) average $V/\sigma$, (iii) average fractional change in $V/\sigma$ between $z = 3$ and $z = 0$, (iv) average local density percentile (see text and Martin et al. (2018a)) , (v) the average fraction of time that the galaxies have spent with the morphology that they have at $z = 0$.

| $\log_{10}(M_\star/M_\odot)$ | $N$ (i) | $V/\sigma$ (ii) | $\Delta V/\sigma$ (iii) | Environment (iv) | time with morphology (v) |
|---|---|---|---|---|---|
| 10.5–11.0 | 2064 | $0.333(0.359)^{+0.078}_{-0.107}$ | $-0.544(-0.629)^{+0.187}_{-0.032}$ | $49.16(48.23)^{+18.17}_{-16.99}$ | $0.498(0.493)^{+0.137}_{-0.156}$ |
| | 6056 | $0.898(0.896)^{+0.111}_{-0.099}$ | $0.110(-0.051)^{+0.313}_{-0.013}$ | $49.84(49.83)^{+17.68}_{-17.94}$ | $0.952(0.987)^{+0.019}_{-0.048}$ |
| 11.0–11.5 | 1189 | $0.268(0.254)^{+0.103}_{-0.088}$ | $-0.627(-0.707)^{+0.183}_{-0.038}$ | $47.51(47.16)^{+16.99}_{-15.45}$ | $0.579(0.573)^{+0.123}_{-0.155}$ |
| | 987 | $0.856(0.812)^{+0.135}_{-0.073}$ | $0.149(-0.110)^{+0.408}_{-0.060}$ | $53.18(55.29)^{+18.43}_{-15.95}$ | $0.897(0.973)^{+0.032}_{-0.103}$ |
| 11.5–12.0 | 299 | $0.193(0.156)^{+0.091}_{-0.050}$ | $-0.566(-0.809)^{+0.305}_{-0.115}$ | $49.68(49.43)^{+17.01}_{-17.36}$ | $0.693(0.720)^{+0.133}_{-0.173}$ |
| | 21 | $0.697(0.675)^{+0.067}_{-0.006}$ | $0.094(-0.023)^{+0.277}_{-0.043}$ | $55.07(58.29)^{+23.87}_{-20.87}$ | $0.650(0.662)^{+0.086}_{-0.221}$ |

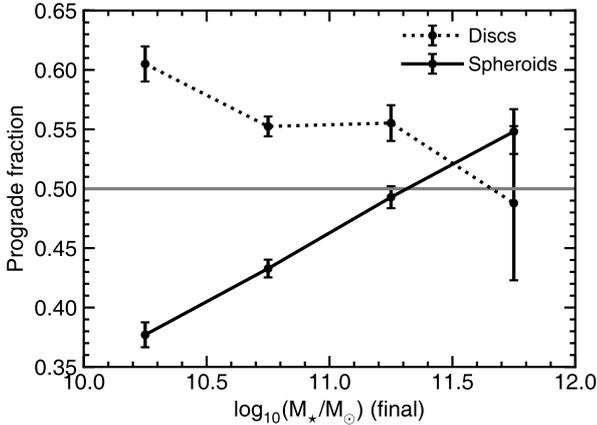

**Figure 7.** The fraction of mergers since $z = 3$ that are prograde in the merger histories of galaxies that are spheroids (solid line) and discs (dotted line) at $z = 0$, as a function of the stellar mass of the galaxy.

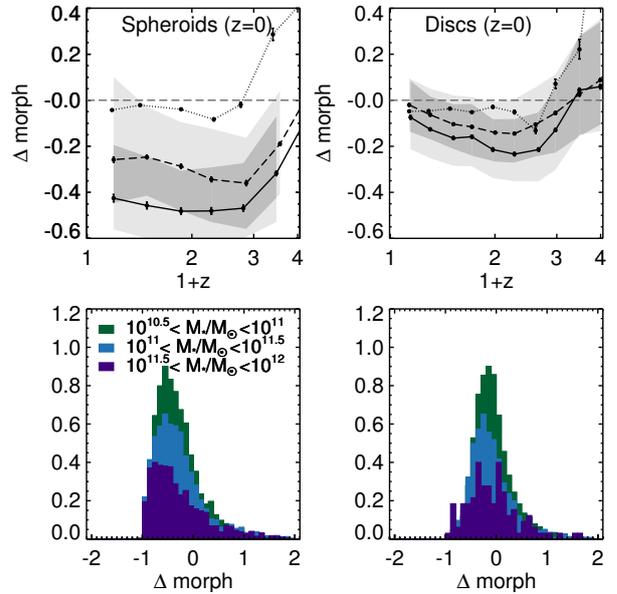

**Figure 8. Top:** Median $\Delta$morph, as a function of redshift, in mergers that involve *progenitors* of galaxies that have stellar masses of $M_\star/M_\odot > 10^{10.5}$ at $z = 0$. The left-hand column shows the redshift evolution of $\Delta$morph for mergers that involve progenitors of spheroids at $z = 0$, while the right-hand column is the corresponding plot for galaxies that are discs at $z = 0$. The filled regions indicate the $\pm 1\sigma$ dispersions and the dark solid and dashed lines indicate the median $\Delta$morph for major and minor mergers respectively. The dotted lines indicate the median $\Delta$morph when these progenitors were not merging. **Bottom:** Histograms showing the corresponding distributions of $\Delta$morph values for major + minor mergers since $z = 3$. Colours indicate the final stellar masses of galaxies at $z = 0$ (see legend).

## 4 THE CUMULATIVE EFFECT OF MERGERS OVER COSMIC TIME

In this section, we investigate the cumulative effect that mergers have on the morphological evolution of galaxies over cosmic time. We discuss the cumulative evolution of $V/\sigma$ as a function of mass and redshift (Section 4.1), present the average merger histories of spheroids and discs (Section 4.2), explore the cumulative effect of mergers on $V/\sigma$ in both spheroids and discs (Section 4.3) and quantify the relative role of major and minor mergers (and other potential processes) in driving the overall evolution of galaxy morphology over cosmic time (Section 4.4).

### 4.1 Galaxy morphology over cosmic time

Figure 9 shows a projection through the Horizon-AGN simulation volume, with the $V/\sigma$ and stellar mass of galaxies shown using the colour and size of the points respectively. At $z = 3$, the galaxy population is relatively homogeneous in a morphological sense - at this epoch, the mean $V/\sigma$ of the progenitors of today's disc and spheroidal galaxies are 0.95 and 0.9 respectively and very few galaxies have gained spheroidal morphologies. This indicates that significant morphological transformation is not yet underway at this redshift. On average, $V/\sigma$ decreases towards the present day, with the most massive galaxies ($M_\star > 10^{11} M_\odot$) dominated by spheroidal morphologies in the local Universe. On the other hand, intermediate mass galaxies ($10^{10.5} < M_\star/M_\odot < 10^{11}$) undergo relatively little morphological change ($V/\sigma$ is reduced by less than 5 per cent between $z = 3$ and today on average) and the population remains dominated by discs at $z = 0$.

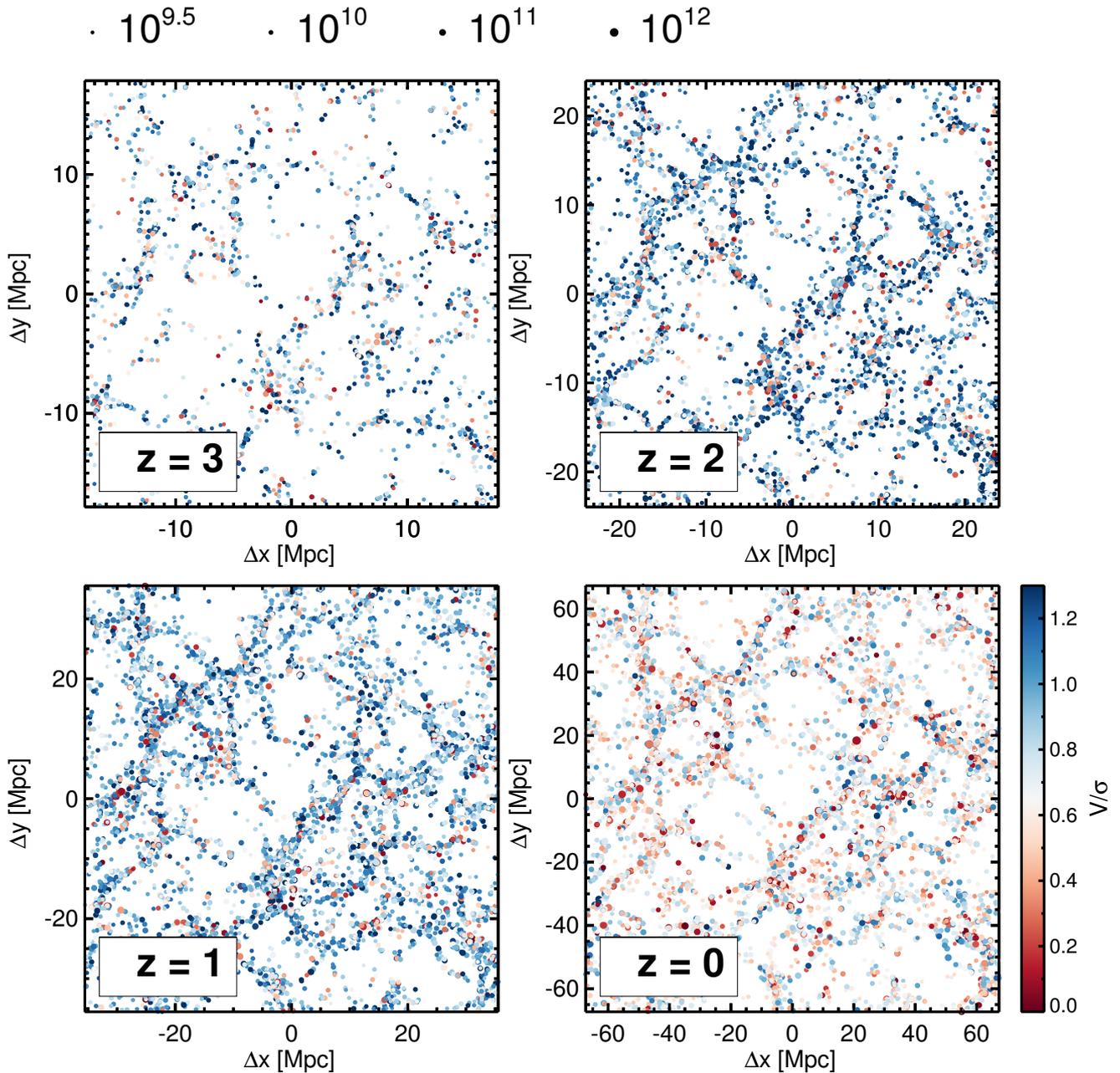

**Figure 9.** A projection through a 20 coMpc slice of the Horizon-AGN simulation volume showing the position (in proper Mpc) and $V/\sigma$ of galaxies with stellar masses greater than $10^{9.5} M_\odot$. Each panel shows the simulation at a different redshift. $V/\sigma$ is represented by the colour of each point (see colour bar). The stellar mass of each galaxy is represented by the size of the symbol, as indicated in the legend above the plots.

Table 1 summarises the average properties of galaxies at $z = 0$ as well as aspects of their morphological evolution. Massive spheroids have undergone significant morphological transformation between $z = 3$ and today (the value of $V/\sigma$ today is at least 50 per cent of the value they had at at $z = 3$, for $M_\star > 10^{10.5} M_\odot$). Lower mass spheroids tend to have attained spheroidal morphology later in their lifetime. Discs on the other hand undergo almost no morphological change over this period. Furthermore, as might be expected, discs are unlikely to have spheroidal morphologies at any point in their lifetime, although, interestingly, the main progen-itors of extremely massive discs, that have $10^{11.5} < M_\star/M_\odot < 10^{12}$ today, spend around a third of their time as spheroids. We return to this point in the next section.

We note that galaxy morphology does not appear to be strongly correlated with galaxy environment (column (iv) of Table 1). Following Martin et al. (2018a), we estimate environment by first ranking each galaxy by their local number density, calculated using an adaptive kernel density estimation method (Breiman et al. 1977). Galaxies are then sorted into density percentiles, so that galaxies in e.g. the 0-10th percentile range represent those in the least dense environments and those in the 90-100$^{th}$ percentile range represent the most dense (we refer readers to Martin et al. 2018a, for more details of this procedure). The average environments of spheroids and discs are reasonably similar, when controlled for stellar mass. As we show later in Section 4.3, environment is only important for the morphological transformation of intermediate mass galaxies in the most extreme environments i.e. clusters (where a minority of the overall galaxy population is found).

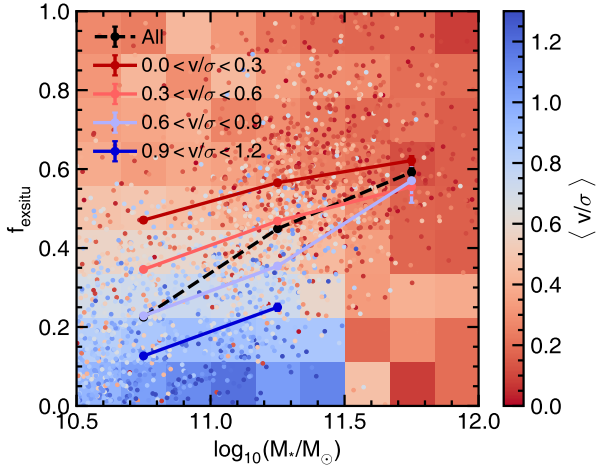

**Figure 10.** 2-D histogram indicating the mean $V/\sigma$ as a function of stellar mass and ex-situ mass fraction ($f_{exsitu}$). Overlaid is a scatter plot of 2000 randomly selected galaxies, and lines indicating the mean ex-situ mass fraction in four $V/\sigma$ bins (see legend) and the whole population (black). $V/\sigma$ values for the histogram and points are indicated by the colour bar.

### 4.2 Average merger histories of spheroids and discs

Table 2 shows the average merger histories of today's spheroids and discs in different mass ranges. On average, today's spheroids have undergone a greater number of major and minor mergers than discs. For example, present day spheroids in the stellar mass range $10^{10.5} < M_\star/\mathrm{M}_\odot < 10^{11}$ have experienced an average of 1.18 (2.30) mergers with mass ratios $> 1 : 4$ (mass ratios $> 1 : 10$), since $z = 3$. This corresponds to 20 (34) per cent of time spent in merging episodes on average. Present-day discs in the same mass range spend less time merging, undergoing 0.73 (1.54) such mergers since $z = 3$, corresponding to 11 (22) per cent of their lifetime since $z = 3$. Similarly, present day spheroids in the stellar mass range $10^{11} < M_\star/\mathrm{M}_\odot < 10^{11.5}$ undergo 1.35 (2.86) mergers, whereas discs undergo 0.95 (2.10) mergers, corresponding to 27 (47) per cent and 17 (34) per cent of their lifetime for spheroids and discs respectively. Spheroids also tend to undergo mergers of larger mass ratios – on average, the largest mass ratio merger undergone by a $10^{10.5} < M_\star/\mathrm{M}_\odot < 10^{11}$ spheroid is 1:2.9, whereas for discs it is 1:3.2. For $10^{11} < M_\star/\mathrm{M}_\odot < 10^{11.5}$ these values are 1:2.8 and 1:3.2 for spheroids and discs respectively.

Figure 10 is a 2-D histogram showing the mean value of $V/\sigma$ (indicated by the colour-bar) as a function of the stellar mass of galaxies at the present day and their ex-situ mass fractions ($f_{exsitu}$), i.e. the fraction of stellar mass directly accreted from other objects via mergers. Coloured points in Figure 10 show the $V/\sigma$ of a randomly selected sample of galaxies. Not unexpectedly, discs dominate the region of parameter space where $f_{exsitu}$ is low ($< 0.3$), except for $M_\star > 10^{11.5}$, because galaxies are typically unable to reach the highest stellar masses through secular growth alone. At low $f_{exsitu}$, stellar mass is not a strong predictor of morphology and, as the blue region does not evolve appreciably towards higher stellar mass. This indicates that stellar mass does not correlate significantly with morphology beyond the trend between stellar mass and ex-situ mass fraction.

Spheroids are the dominant morphological type at high $f_{exsitu}$, reflecting the important role that mergers play in morphological transformation. More massive galaxies are, therefore, more spheroidal on average, because mergers are the primary means by which they grow their stellar mass. However, some interesting subpopulations become apparent in this figure - at high $f_{exsitu}$, a significant population of discs (high $V/\sigma$) remains and there is a population of massive, low $f_{exsitu}$ slowly-rotating spheroids. For example, 13 per cent of galaxies with exsitu mass fractions greater than 0.8 still have disc-like morphologies, and 6 per cent of galaxies with exsitu mass fractions less than 0.2 have spheroidal morphologies. Although these populations are in the minority, this reflects the diversity of formation channels for discs and spheroids. We will study the formation of these sub-populations in detail in two forthcoming papers. These will show that extremely massive discs (which all have high ex-situ mass fractions) are the result of very recent disc rejuvenation from gas-rich mergers (Jackson et al. in prep), while slowly-rotating spheroids with low ex-situ mass fractions are typically the result of single minor-merger events where the orbits of the satellites at coalescence are close to being co-planar with the disc of the more massive merger progenitor (Jackson et al. in prep).

### 4.3 Cumulative impact of mergers on galaxy morphology over cosmic time

In Figure 11, we study the evolution, over cosmic time, of the mean $V/\sigma$ of galaxies, and the contribution of mergers to the overall morphological transformation in galaxies that are spheroids (left-hand column) and discs (right-hand column) at the present day. As noted in Section 2.2 above, the accuracy of such a cumulative analysis is sensitive to the completeness of the merger history. Hence, we restrict this analysis to galaxies in the regime where completeness is high ($M_\star > 10^{10.5}$ M$_\odot$).

In each panel, we show the average change in $V/\sigma$ of the population in question (dark solid line), the average change contributed just by major mergers (light solid line) and the average change contributed by both major and minor mergers (dashed line). For example, if the mean evolution of $V/\sigma$ for a galaxy population, shown by the dark solid line, overlaps perfectly with that due to mergers (i.e. the dashed line), then mergers are responsible for all the morphological transformation in these galaxies. Similarly, if the mean $V/\sigma$ at the present day is higher than what would be expected due to mergers (i.e. the dark solid line is above the dashed line at $z \sim 0$), then there must be other processes (e.g. accretion) that are responsible for spinning the galaxies up over their lifetimes.

For spheroids that have stellar masses greater than $10^{10.5}$M$_\odot$ at $z = 0$, major and minor mergers (i.e. mergers with mass ratios greater than $1 : 10$) together explain essentially all the morphological evolution. However, it is important to note that the morphological evolution is not induced by major mergers alone since, if that had been the case, the dark solid lines would have overlapped with the lighter solid lines (which correspond to just major mergers). Indeed, the overall change in $V/\sigma$ of such massive spheroids is larger than what can be produced by major mergers alone, indicating that *minor mergers have a significant role to play in the transformation of morphology (disc to spheroid) over cosmic time*.

**Table 2.** Mean (median) merger histories of spheroids (white rows) and discs (shaded rows) for galaxies at $z = 0$. $\pm$ indicates the $1\sigma$ dispersion. For columns (iii) and (iv) we report the mean and standard deviation only. (i) The average redshift at which the largest mass ratio merger occurred, (ii) the average mass ratio of the largest mass ratio merger, (iii) the average number of major mergers undergone, (iv) the average number of minor mergers undergone, (v) the average ex-situ mass fraction at $z = 0$.

| $\log_{10}(M_*/M_\odot)$ | $z$ of largest (i) | largest mass ratio (ii) | # major (iii) | # minor (iv) | ex situ mass fraction (v) |
|---|---|---|---|---|---|
| 10.5–11.0 | $1.224(1.092)^{+0.509}_{-0.341}$ | $2.864(2.485)^{+0.996}_{-0.238}$ | $1.178 \pm 0.921$ | $1.118 \pm 1.063$ | $0.412(0.401)^{+0.089}_{-0.070}$ |
| | $1.169(1.027)^{+0.614}_{-0.396}$ | $3.249(2.719)^{+1.413}_{-0.444}$ | $0.726 \pm 0.803$ | $0.816 \pm 0.872$ | $0.201(0.176)^{+0.091}_{-0.050}$ |
| 11.0–11.5 | $1.203(1.027)^{+0.488}_{-0.322}$ | $2.815(2.367)^{+0.978}_{-0.284}$ | $1.355 \pm 1.032$ | $1.504 \pm 1.185$ | $0.542(0.533)^{+0.078}_{-0.061}$ |
| | $1.121(0.968)^{+0.601}_{-0.364}$ | $3.244(2.679)^{+1.291}_{-0.419}$ | $0.946 \pm 0.981$ | $1.150 \pm 1.052$ | $0.336(0.321)^{+0.093}_{-0.059}$ |
| 11.5–12.0 | $1.235(1.092)^{+0.520}_{-0.369}$ | $3.076(2.523)^{+1.141}_{-0.282}$ | $1.217 \pm 1.114$ | $1.548 \pm 1.246$ | $0.609(0.597)^{+0.081}_{-0.067}$ |
| | $0.901(0.632)^{+0.573}_{-0.538}$ | $2.210(2.099)^{+0.538}_{-0.128}$ | $1.238 \pm 0.921$ | $1.714 \pm 1.201$ | $0.536(0.555)^{+0.070}_{-0.119}$ |

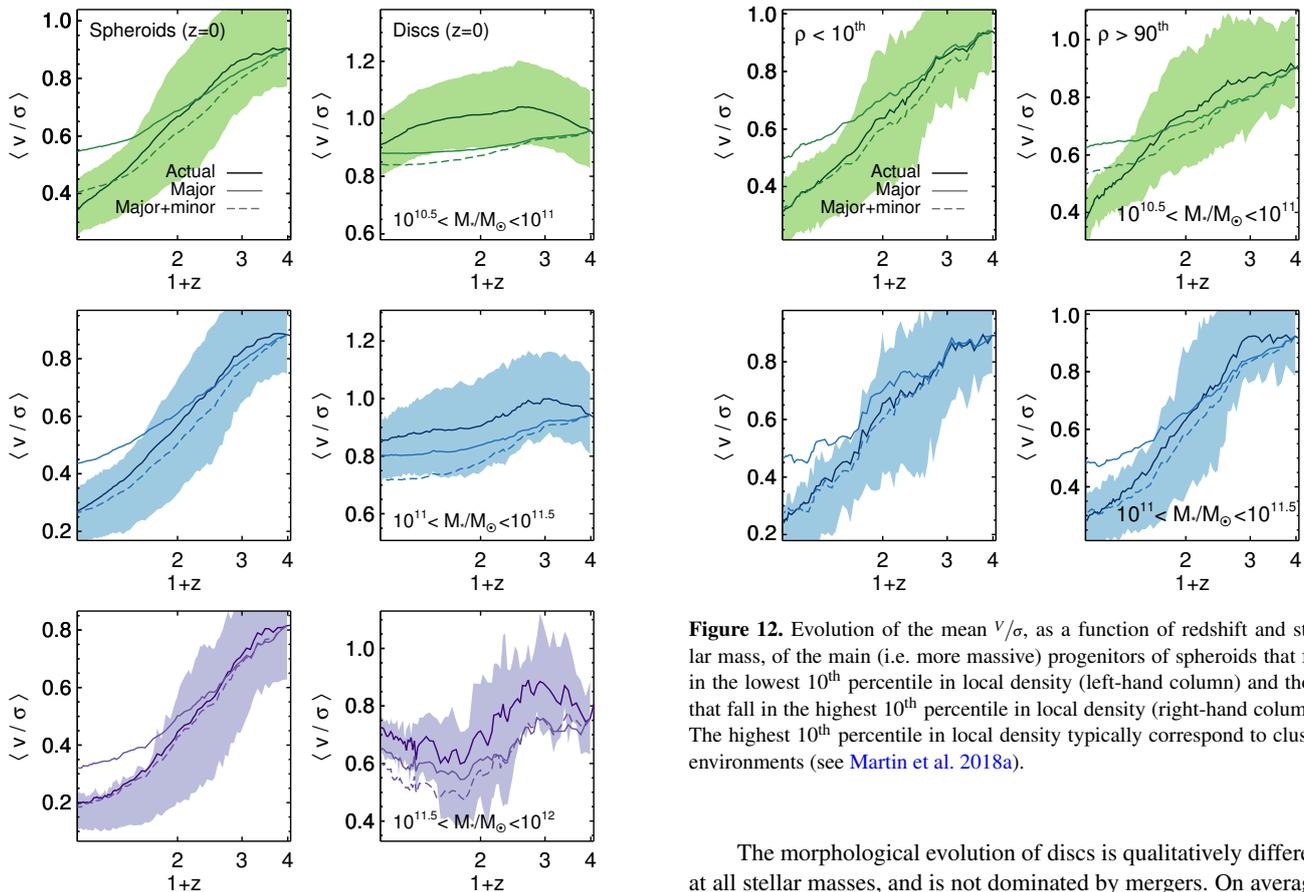

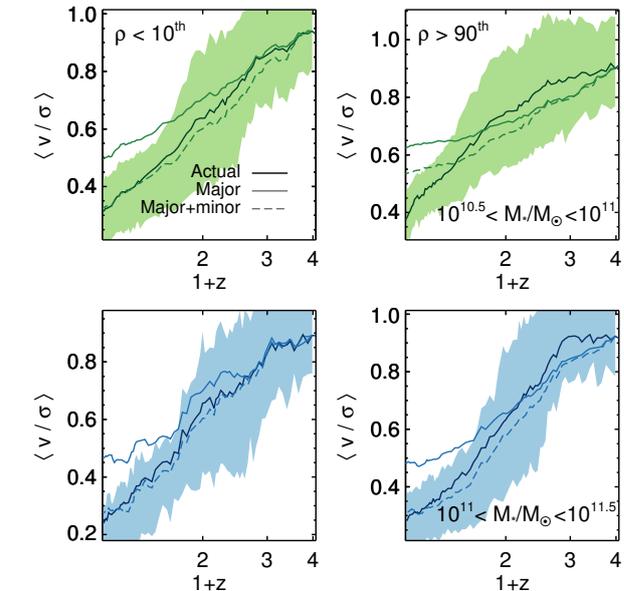

**Figure 11.** Evolution of the mean $v/\sigma$, as a function of redshift and stellar mass, of the main (i.e. more massive) progenitors of spheroids (left-hand column) and discs (right-hand column). The dark solid line indicates the actual $v/\sigma$ evolution, the light solid line indicates the $v/\sigma$ evolution due to major mergers alone and the dashed line shows the $v/\sigma$ evolution due to major + minor mergers.

Indeed, after $z \sim 1$ (where the light solid and dashed lines start diverging), minor mergers are responsible for the majority of the morphological transformation in spheroids. It is worth noting here that, since all the morphological transformation can be accounted for by major and minor mergers, other processes, such as fly-bys, or the formation of low $v/\sigma$ stars from low-angular momentum gas (e.g. via feeding from counter-rotating filaments where the net angular momentum is close to zero, e.g. Danovich et al. (2015) are unlikely to be significant drivers of morphological transformation over cosmic time.

**Figure 12.** Evolution of the mean $v/\sigma$, as a function of redshift and stellar mass, of the main (i.e. more massive) progenitors of spheroids that fall in the lowest $10^{th}$ percentile in local density (left-hand column) and those that fall in the highest $10^{th}$ percentile in local density (right-hand column). The highest $10^{th}$ percentile in local density typically correspond to cluster environments (see Martin et al. 2018a).

The morphological evolution of discs is qualitatively different at all stellar masses, and is not dominated by mergers. On average, the $v/\sigma$ values of disc galaxies today are higher than they would be if their evolution was being driven by mergers. This indicates that discs differ from spheroids, in the sense that they are consistently spun up by gas accretion at all epochs. This accretion counteracts the decrease in $v/\sigma$ due to merging, especially at high redshift when the Universe is more gas rich.

It is important to note that, while mergers explain the majority of the morphological evolution that leads to the formation of spheroids, the morphological evolution of the disc population *cannot* be entirely explained by a *lack* of mergers. As described above in Table 2, while spheroids do tend to have more mergers than discs at a given stellar mass, the average merger histories of the two morphological classes are not too dissimilar. This is not surprising, since the merger history of a galaxy is expected to be a strong function of its stellar mass (e.g. Stewart et al. 2008; Rodriguez-Gomez et al. 2015). At a given stellar mass, therefore, spheroids do not undergo many more mergers than discs, nor do they undergo mergers of appreciably higher mass ratios (Table 2). However, the mergers that discs do undergo clearly do not produce the same morphological change as that seen in mergers that involve spheroids. For example, mergers with mass ratios $> 1 : 10$ produce a mean fractional change $((\Delta^V/\sigma)/(^V/\sigma)_{z=3})$ of -0.72 for spheroids of stellar mass $10^{11} < M_\star/M_\odot < 10^{11.5}$ between $z = 3$ and $z = 0$, yet only produce a fractional change of -0.12 in discs, where $\Delta^V/\sigma$ is the $^V/\sigma$ at $z = 0$ subtracted from that at $z = 3$.

The explanation lies largely in the actual properties of the galaxies and mergers themselves. As described in Section 4.1, discs have higher gas fractions, for a given stellar mass, than spheroids. Since gas rich mergers have a higher likelihood of producing $\Delta$morph $> 0$ (Figure 3), any mergers they undergo tend not to significantly decrease $^V/\sigma$, as new stars formed from the residual gas act to counteract some of the morphological transformation, by adding to the rotational component of the system. Disc rejuvenation, especially at early epochs, can also be assisted by cosmological accretion, which enables galaxies to re-acquire cold gas and reform discs, at least until a critical mass of $\sim 10^{10.5}$ M$_\odot$ (e.g. Welker et al. 2017). Finally, as we have shown in Section 3.3, the spin of the satellite and the orbital configuration of the merger, relative to the spin of the more massive progenitor, plays a role in determining the properties of the remnant. On average, present-day discs have undergone more prograde mergers, whereas their spheroidal counterparts of comparable stellar mass have undergone more retrograde mergers. This also plays a part in preserving the morphological properties of discs.

It is worth noting the fraction of time galaxies are considered to be in a merging phase, given our chosen timescale (2 Gyr). Since more massive galaxies undergo a greater number of mergers, they spend more of their time merging on average. Galaxies in the stellar mass range $10^{11.5} < M_\star/M_\odot < 10^{12}$ undergo major and major+minor mergers for 32 and 55 per cent of their lifetimes respectively. For galaxies with stellar masses in the range $10^{10.5} < M_\star/M_\odot < 10^{11}$, the corresponding values are 13 and 25 per cent. Importantly, our choice of a 2 Gyr timescale does not result in a scenario where most of galaxy's evolution takes place during merging episodes, thereby underestimating the role of secular evolution.

We complete this section by exploring, in Figure 12, the role of environment in producing morphological transformation, by considering the $^V/\sigma$ evolution of spheroid progenitors in the densest and least dense environments. While the morphological transformation of the most isolated galaxies (those in the bottom $10^{th}$ percentile of local number density) is entirely accounted for by major and minor mergers, other processes must be invoked in order to explain the evolution of morphology in the densest environments (those in the top $10^{th}$ percentile, which correspond to clusters, see Martin et al. (2018a)). Mergers account for around two-thirds of the morphological change in intermediate-mass galaxies ($10^{10.5} < M_\star/M_\odot < 10^{11}$) in the densest environments, while the remaining one-third may be due to other processes e.g. harassment. Our result is consistent with the results of Choi & Yi (2017), who find that mechanisms other than mergers contribute to the morphological transformation of such intermediate-mass galaxies in cluster environments. We note that, unlike the simulations of Choi & Yi (2017), rich cluster environments are rare in the cosmological Horizon-AGN volume (there is only one rich Coma-like cluster in the entire volume). Thus, the Horizon-AGN volume is skewed towards low-density environments, where environmental processes are, by definition, less important.

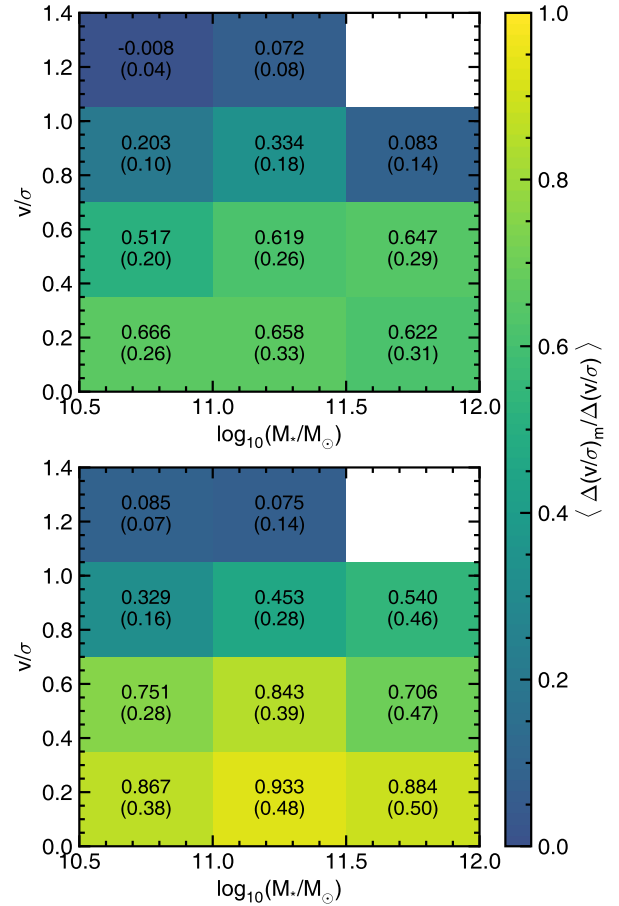

**Figure 13.** 2-D histograms showing the mean fraction of morphological change that occurs during merging episodes, as a function of the stellar mass and $^V/\sigma$ of the remnant at $z = 0$. This mean fraction is indicated by the colour-bar and also shown in the centre of each bin. The bracketed value indicates the ex-situ stellar mass fraction of the remnant. **Top**: The mean fraction of morphological change and ex-situ mass fractions produced by major mergers. **Bottom**: The mean fraction of morphological change and ex-situ mass fractions produced by major + minor mergers.

### 4.4 The fraction of global morphological transformation triggered by major and minor mergers

We complete our study by quantifying the fraction of overall morphological change in massive galaxies, since $z = 3$, that is attributable to mergers. We use the following definition for the fraction of the morphological transformation that is driven by mergers:

$$f = \Delta(^V/\sigma)_m / \Delta(^V/\sigma) , \qquad (6)$$

where $\Delta(^V/\sigma)_m$ is the total change in $^V/\sigma$ during merging episodes since $z = 3$ and $\Delta(^V/\sigma)$ is the total change in $^V/\sigma$ since $z = 3$. By this definition, $f$ can be positive or negative (in cases where the $^V/\sigma$ of galaxies increases between $z = 3$ and today). It may also exceed 1 in cases where the $\Delta(^V/\sigma)_m$ is larger than $\Delta(^V/\sigma)$. In Figure 13 we present 2-D histograms that show the mean value of $f$, as a function of stellar mass and $^V/\sigma$, for major mergers (mass ratios $> 1 : 4$; top panel) and major + minor mergers (mass ratios $> 1 : 10$; bottom panel). Values in brackets indicate the mean fraction of ex-situ stellar mass that is accreted in mergers with these mass ratios in each bin. The most spheroidal galaxies are those with the highest values of $f$. Of the morphological transformation produced by mergers, minor mergers are responsible for a significant minority (around 30 per cent) over cosmic time. And, as noted in the previous section, the role of minor mergers becomes dominant at later epochs ($z < 1$), where they drive almost all the morphological transformation in the spheroid population.

Mergers are the dominant means of morphological transformation in high mass spheroids ($M_\star > 10^{10.5} M_\odot$), where they are responsible for at least 75 per cent of the change. The vast majority (90 per cent) of the morphological change in the most spheroidal galaxies ($V/\sigma < 0.3$) is the result of mergers. In discs with significant dispersion-dominated components, a large proportion of morphological change is a result of mergers ($\sim 40$ per cent), particularly minor mergers. Our results are consistent with those of Choi et al. (2018), who find that the spin change since $z = 1$ in 94 per cent of massive central early-type galaxies is dominated by mergers.

While discs typically have non-negligible ex-situ mass fractions, a larger fraction of this mass is derived from minor mergers (averaged over all masses, around 70 per cent of ex-situ mass in spheroids is derived from major mergers, compared to $50 - 60$ per cent for discs). Minor and major mergers do not play a significant role either in the morphological transformation or mass assembly of the most rotationally-supported discs ($V/\sigma > 1$). Taken together, major and minor mergers account for only $\sim 10$ per cent of the stellar mass of such discs and are responsible for less than 10 per cent of their morphological transformation. What little morphological transformation takes place in these galaxies is instead dominated by spin up due to accretion of gas.

## 5 CONCLUSIONS

We have used Horizon-AGN, a cosmological hydrodynamical simulation, to study the processes that drive morphological transformation across cosmic time. In particular, we have (1) studied the average merger histories of discs and spheroids over cosmic time, (2) quantified the magnitude of the morphological change (i.e. disc to spheroid) that is imparted by major and minor mergers as a function of redshift and stellar mass, (3) explored the effect of gas fraction on these morphological changes, (4) studied the effect of orbital configuration in determining the properties of merger remnants and (5) quantified the overall contribution of major/minor mergers and other processes to the creation of spheroidal galaxies. Our key conclusions are as follows:

- *The morphological evolution of spheroids with stellar masses greater than $10^{10.5} M_\odot$ at $z = 0$ can be largely explained by relatively high-mass-ratio ($> 1 : 10$) mergers.* Essentially all of the morphological evolution in galaxies that are spheroids at $z = 0$ took place in mergers with mass ratios greater than $1 : 10$. However, major mergers (mass ratios $> 1 : 4$) *alone* are not sufficient. Around a third of the overall morphological change in massive spheroids is driven by minor mergers (mass ratios between $1 : 4$ and $1 : 10$). Furthermore, minor mergers become the dominant channel for morphological change at late epochs, driving the bulk of the morphological transformation in spheroid progenitors at $z < 1$. Finally, across the *general* galaxy population (i.e. across all masses and environments), other processes, such as fly-bys, harassment or the formation of low $V/\sigma$ stars from direct accretion of low-angular momentum gas, are relatively insignificant in transforming galaxy morphology, since all the morphological change in massive spheroids is accounted for by mergers with mass ratios greater than $1 : 10$.

- *In clusters, environmental processes like harassment do play a role in morphological transformation for intermediate mass galaxies ($10^{10.5} < M_\star/M_\odot < 10^{11}$).* Around a third of the morphological transformation of $10^{10.5} < M_\star/M_\odot < 10^{11}$ spheroids in the *densest* environments is a result of processes other than mergers. However, since the vast majority of galaxies do not reside in clusters, almost all morphological transformation in the general population of massive spheroids can be explained by mergers alone.

- *The outcome of a merger is strongly influenced by the gas fraction of the merging pair.* Mergers involving spheroids with higher gas fractions are more likely to produce remnants with increased $V/\sigma$. For example, at $z \sim 0$, gas poor major mergers ($f_{gas} < 0.2$) typically produce fractional morphological changes ($\Delta$ morph) of -0.1 in spheroids, whereas gas rich major mergers ($f_{gas} > 0.3$) typically produce significantly more positive changes of 0.38 over the course of each merger for galaxies in the mass range $10^{10.5} < M_\star/M_\odot < 10^{11}$.

- *The orbital configuration of a merger has a measurable impact on the properties of the remnant.* Mergers that are prograde (i.e. where the spin of the satellite and the orbital angular momentum are aligned with the spin of the more massive galaxy) typically produce smaller decreases in the spin of discs (i.e. milder morphological transformation) than their retrograde counterparts. while both types of events produce similar morphological changes at $z > 2$, the average change due to retrograde mergers is around twice as large as that due to their prograde counterparts at $z \sim 0$.

- *Spin up due to cosmological accretion is an important effect, especially at early epochs.* In the early universe ($z > 2$), stellar mass forms in discs more rapidly than it can be removed by mergers. At later times, gas-rich minor mergers and existing reservoirs of gas in the halo become increasingly important for spinning up galaxies as the effectiveness of cosmological accretion declines.

- *On average, spheroids have undergone more mergers (and mergers with higher mass ratios) since $z = 3$ compared to discs of equivalent stellar mass (Table 2).* The fraction of stellar mass formed ex-situ (i.e. accreted directly via mergers) is around 1.5–2 times higher in massive spheroids. In addition, the average fractional morphological transformation ($\Delta$morph) induced per merger is around a factor of 2 larger in disc progenitors compared to in spheroid progenitors.

- *However, the survival of discs to $z = 0$ cannot be explained solely by a lack of mergers.* Although the progenitors of discs undergo fewer mergers than equivalent mass spheroids, the disparity between their merger histories is not large enough to account for their relative lack of morphological evolution. The discrepancy instead stems from the properties of the mergers themselves. Mergers involving the progenitors of today's discs tend to be more gas-rich, which promotes disc rejuvenation subsequent to a merger. In addition, discs typically undergo a greater fraction of prograde mergers (compared to spheroids of a similar stellar mass) which induce milder morphological transformation and thus assist in disc survival.

Finally, while we have quantified the creation of spheroids in a broad sense in this paper, it is worth recalling some interesting sub-populations identified in this study, which will be the subject of forthcoming work. First is the existence of a small population of very massive discs ($M_\star/M_\odot > 10^{11.5}$), which have rich merger histories like their spheroidal counterparts (see Table 2) and, as a result, high ex-situ mass fractions. Given that mergers are significant drivers of morphological transformation, a key question is: why do such extremely massive discs exist at all? As we will show in a forthcoming paper (Jackson et al. in prep), the existence of these disks is due to very recent disk rejuvenation via gas-rich mergers. Equally interesting is the population of massive, slowly-rotating (spheroidal) galaxies with low ex-situ mass fractions, which indicate that these systems have not undergone many mergers, yet are not rotationally-supported as as one might expect of a galaxy whose assembly history is largely merger-free. As we will demonstrate in another paper (Jackson et al. in prep), these systems are created by a minor-merger, where the orbit of the satellite at coalescence is almost co-planar with the disc of the more massive merger progenitor (Jackson et al. in prep).

## ACKNOWLEDGEMENTS

We are grateful to the anonymous referee for many constructive comments that improved the quality of this paper. GM acknowledges support from the Science and Technology Facilities Council [ST/N504105/1]. SK acknowledges a Senior Research Fellowship from Worcester College Oxford. JD acknowledges funding support from Adrian Beecroft, the Oxford Martin School and the STFC. This research has used the DiRAC facility, jointly funded by the STFC and the Large Facilities Capital Fund of BIS, and has been partially supported by grant Spin(e) ANR-13-BS05-0005 of the French ANR. This work was granted access to the HPC resources of CINES made by the allocations 2013047012, 2014047012 and 2015047012 made by GENCI. This work is part of the Horizon-UK project.

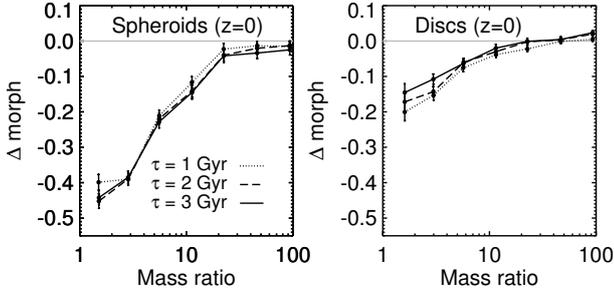

**Figure A1.** The median fractional change in $V/\sigma$ ($\Delta$ morph) over the course of a merger, as a function of merger mass ratio, for mergers at $z < 0.5$ where the more massive galaxy has a stellar mass of at least $10^{10.5}$ M$_\odot$ at the time of the merger. Note that, in order to control for the morphological change that is due to processes other than mergers, we have calculated the average fractional change in $V/\sigma$ for non-merging systems over the same time period and subtracted this value from $\Delta$morph. Dotted, dashed and solid black lines indicate the median value of $\Delta$ morph when merger timescales of 1 Gyr, 2 Gyrs or 3 Gyrs are used respectively. The left-hand panel shows the result for mergers involving the progenitors of today's spheroids, while the right-hand panel shows the same for the progenitors of today's discs.

## APPENDIX A: ROBUSTNESS OF RESULTS WITH RESPECT TO CHOICE OF MERGER TIMESCALE AND MASS RATIO THRESHOLD

In this section, we consider the robustness of our results to both changes in the assumed merger timescale and the range of mass ratios considered in this study. Figure A1 shows the average fractional morphological change ($\Delta$ morph) for the progenitors of today's spheroids and discs as a function of merger mass ratio. This is shown both for our adopted merger timescale of $\tau = 2$ Gyrs, as well as for a shorter 1 Gyr window ($t = \pm 0.5$ Gyrs centred around coalescence) and a longer 3 Gyr window ($t = \pm 1.5$ Gyrs). Black lines indicate the median morphological change for all mergers since $z = 0.5$ once the average fractional morphological change in the non-merging population over the same timescale has been subtracted. We restrict the plot to $z < 0.5$ in order to exclude, as much as possible, the effect of cosmological accretion and because of the dependence of $\Delta$ morph on redshift. However, the curves presented in Figure A1 reach an asymptote at similar mass ratios regardless of the redshift range chosen.

We note first that an important consideration when selecting a timescale is to use a value that encompasses the merger event completely. In particular, choosing a timescale that is too short will lead to spurious results, because the merger remnant will not have relaxed at the point at which it is observed. For mergers that we are interested in studying here (mass ratios around 1:10 or greater), merger timescales are $\sim 2$ Gyr (e.g. Jiang et al. 2008; Kaviraj et al. 2011). Figure A1 shows excellent convergence in the $\Delta$morph values for timescales of 2 and 3 Gyrs. It also shows that for these timescales mergers with very low mass ratios (i.e. less than $\sim 1$ : 10) do not induce large changes in galaxy morphology (we reinforce this point in our discussion of Figure A3 below). For such low mass ratios, the morphological change during mergers for either spheroid and disc progenitors is not appreciably larger than the morphological change observed in the non-merging population over the same time period.

We also show the values of $\Delta$morph for a 1 Gyr timescale, which, as noted above, may be too short for the mass ratios of interest in this study. For the most part, the 1 Gyr timescale gives the same result as the 2 Gyr and 3 Gyr timescales, although they diverge slightly at large mass ratios. Visual inspection of the stellar mass distribution of a subset of galaxies undergoing mergers indicates that, 500 Myrs after coalescence, a large number of these systems have obvious asymmetries (i.e. the merger is not yet complete and the remnant has not relaxed), which will produce spurious values of $V/\sigma$. Thus, while timescales need to be short enough that significant amounts of morphological change are not missed or erroneously counted as merger-induced, it is important that the timescale is long enough that it probes the full duration of the merger.

Figure A2 shows a version of the left-hand panels of Figure 11 (which describe the cumulative change in $V/\sigma$ in the progenitors of today's spheroids) using different merger timescales. We show both the timescale we have adopted for this study (2 Gyr), as well as timescales of 3 Gyr and 1 Gyr (although note from the arguments above that a 1 Gyr timescale is inappropriately short for the mass ratios probed in this study). We find that reducing the merger timescale to 1 Gyr or increasing it to 3 Gyrs introduces only minimal change. We find good convergence between the lines showing the change in average $V/\sigma$ due to mergers for the 2 Gyr and 3 Gyr timescales. The absolute difference between these lines is less than 0.05 for both major and major+minor mergers at all redshifts and in every mass bin). While there is good convergence, the difference between the 1 Gyr and 2 Gyr lines is slightly larger for the highest mass bin. The lines showing the change in average $V/\sigma$ due to mergers, when calculated using 1 Gyr and 2 Gyr timescales, differ by 0.1 by $z = 0$, for $M_\star/M_\odot > 10^{11}$). Nevertheless, the choice of a 3 Gyr or even a 1 Gyr timescale does not qualitatively affect our results.

For large stellar masses ($10^{11} < M_\star/M_\odot < 10^{12}$), where the $V/\sigma$ evolution attributed to mergers is most sensitive to the timescale chosen, the fraction of morphological transformation that we attribute to mergers (Equation 6, Figure 13) remains essentially unchanged. In the case of spheroids, there is a typical absolute increase of only 0.01 in the fraction of morphological change that we attribute to mergers, when a 3 Gyr timescale is used compared to a 2 Gyr timescale. When decreasing the timescale to 1 Gyr, there is a typical absolute decrease of 0.08 in terms of the fraction of morphological change attributed to mergers when compared with a 2 Gyr timescale. The values are similar in the case of discs in the same mass range. There is a typical absolute increase of 0.03 for a 3 Gyr timescale and typical absolute decrease of 0.06 for a 1 Gyr timescale. The exact choice of timescale (1 - 3 Gyrs) is, therefore, relatively unimportant.

Finally, in Figure A3 we show a version of the left-hand panels of Figure 11, now with an additional mass-ratio cut of 1:20. In line with the results of Figure A1, we find that considering mass ratios less than 1:10 does not change the results of this study. In other words, mass ratios less than 1:10 produce negligible amounts of morphological transformation.

This paper has been typeset from a TeX/LaTeX file prepared by the author.

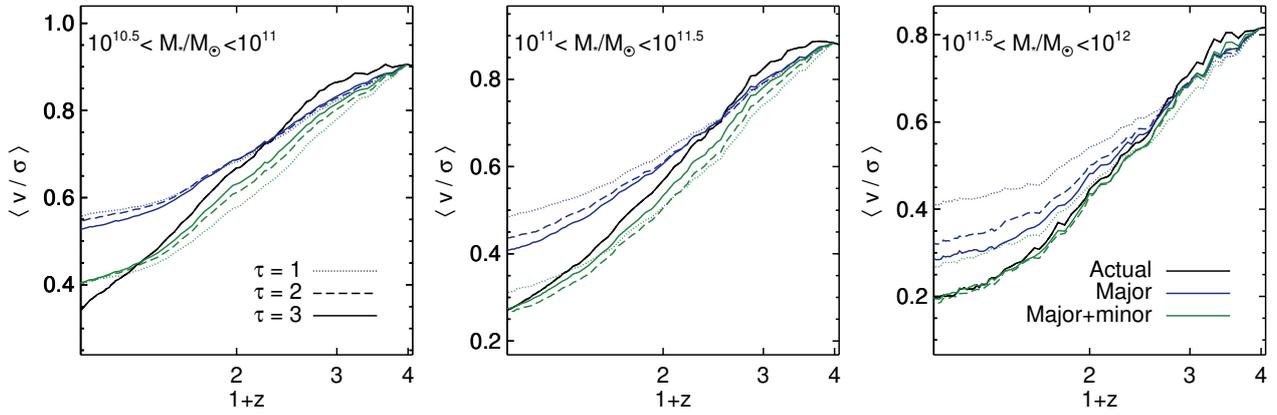

**Figure A2.** Left-hand panel of Figure 11, now plotted for merger timescales of 1 Gyr, 2 Gyrs or 3 Gyrs, indicated by dotted, dashed and solid lines respectively. Blue lines indicate morphological transformation due to major mergers alone, while green lines indicate morphological change that is due to major+minor mergers.

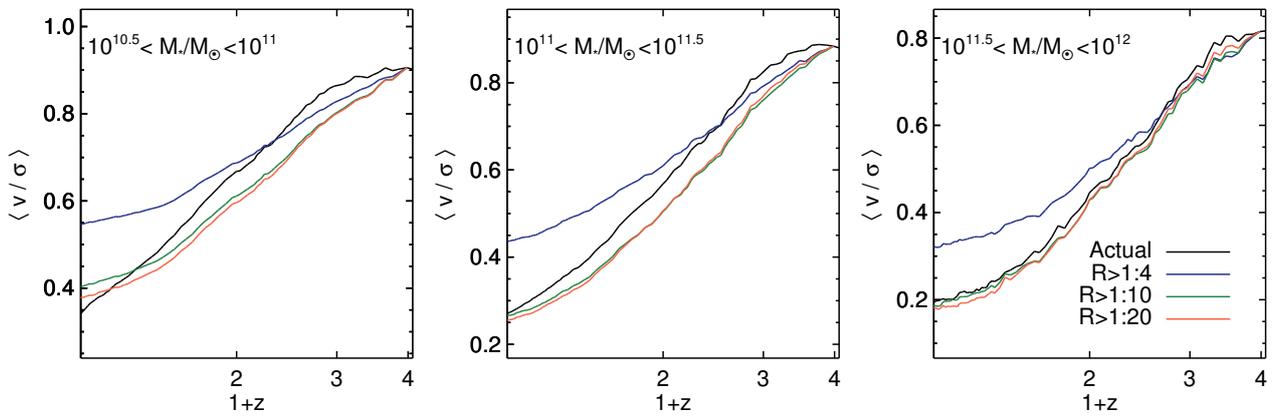

**Figure A3.** Left-hand panel of Figure 11 plotted for additional mass ratio cuts. *R* represents the stellar mass ratio of mergers. Blue lines indicate morphological transformation due to major mergers alone (R> 1 : 4) and green lines indicate morphological change that is due to major+minor mergers (R> 1 : 10). Red lines indicate the morphological change due all mergers down to a mass ratio of 1:20.